\documentclass[12pt,a4paper]{article}
\usepackage[latin1]{inputenc}
\usepackage{amsmath,bm}
\usepackage{amsfonts}
\usepackage{amssymb}
\usepackage{float}
\usepackage{cite}
\usepackage{breqn}
\usepackage{xcolor, soul}
\sethlcolor{yellow}
\makeatletter
\newcommand{\mathleft}{\@fleqntrue\@mathmargin0pt}
\newcommand{\mathcenter}{\@fleqnfalse}
\makeatother
\newcommand*{\SavedEqref}{}
\let\SavedEqref\eqref
\renewcommand*{\eqref}[1]{%
  \begingroup
    \hypersetup{
      linkcolor=linkequation,
      linkbordercolor=linkequation,
    }%
    \SavedEqref{#1}%
  \endgroup
}
\usepackage{graphicx}
\DeclareGraphicsExtensions{.pdf,.png,.jpg}
\usepackage[a4paper, width = 160mm, top = 30mm, bottom = 25mm, 
bindingoffset = 0mm]{geometry}
\def\beq{\begin{equation}}
\def\eeq{\end{equation}}
\def\bea{\begin{eqnarray}}
\def\eea{\end{eqnarray}}

\begin{document}
\begin{center}
  {\Large \bf Tunneling from general Smith-Volterra-Cantor potential }
\vspace{1.3cm}

{\sf Vibhav Narayan Singh\footnote[1]{e-mail address:\ \ vibhavn.singh13@bhu.ac.in,\hspace{0.05cm} vibhav.ecc123@gmail.com},
Mohammad Umar\footnote[2]{e-mail address:\ \ pha212475@iitd.ac.in},
Mohammad Hasan\footnote[3]{e-mail address:\ \ mhasan@isro.gov.in,\hspace{0.05cm} mohammadhasan786@gmail.com},\\
Bhabani Prasad Mandal\footnote[4]{e-mail address:\ \ bhabani.mandal@gmail.com,\hspace{0.05cm} bhabani@bhu.ac.in}}

\bigskip

{\em $^{1,4}$Department of Physics,
Banaras Hindu University,
Varanasi-221005, INDIA. \\ 
$^{2}$Indian Institute of Technology, Delhi-110016, INDIA. \\
$^{3}$Indian Space Research Organisation,
Bangalore-560094, INDIA. \\}

\bigskip	
	\noindent {\bf Abstract}
	
\end{center}

We study the tunneling problem from general Smith-Volterra-Cantor (SVC) potential of finite length $L$ characterized by the scaling parameter $\rho$ and stage $G$. We show that the SVC($\rho$) potential of stage $G$ is the special case of super periodic potential (SPP) of order $G$. By using SPP formalism developed by us earlier, we provide the close form expression of tunneling probability $T_{G}(k)$ with the help of $q$-Pochhammer symbol. The profile of $T_{G}(k)$ with wave vector $k$ is found to saturate with increasing stage $G$. Very sharp transmission resonances are found to occur from this system which may find applications in the design of sharp transmission filters.

\medskip
\vspace{1in}
\newpage

\section{Introduction}

The tunneling of a quantum particle is one of the most fundamental and earliest studied problem of quantum mechanics which started in the year $1928$ \cite{ nordheim1928, gurney1928}. Since then the quantum mechanical tunneling has long been studied by several authors \cite{condon, wigner_1955, david_bohm_1951, book1, book2}. The propagation of matter waves through potential distributions have been extensively studied due to their various applications in theory, experiments as well as for practical applications. Various types of potentials have been studied extensively and the range of solvable potentials is extended by various methods \cite{book1, book2}. The transmission of electromagnetic and quantum waves through potentials having fractal distributions has attracted much attention due to the sharp transmissions, wave localization and self-similarity of physical properties \cite {mandelbrot,feder,wen,shalaev1,shalaev2,shalaev3,takeda,chuprikov2000,miyamoto, cantor_graphene}. Some of these phenomena have also been experimentally observed \cite{takeda, miyamoto}.

Fractals are geometric objects that are self-similar and homogeneous. Mathematically, fractals are self-similar geometric structures that are obtained by a basic mathematical operation on a geometrical object. The mathematical operation is called `generator' while the geometrical object on which this operation is performed is called `initiator'. The process of basic operation is repeated on multiple levels. At each level, a geometrical object having sub-units are created that resembles the structure of the whole object. Due to the continuous nature of the division of real numbers, this results in the resemblance of the sub-units to the whole object i.e. self-similar property hold at all scales for fractals.  Fractal patterns also occur in nature and usually describe naturally occurring fragmented and irregular structures \cite{mandelbrot, voss, hurd}. However, in nature, the self-similarity doesn't hold at all scales and in general, there exists an upper and lower limit within which the self-similarity applies. 
	
One of the simplest scattering problems in the fractal system is one-dimensional scattering by a Cantor fractal potential. Cantor fractal potentials have been extensively studied in quantum mechanics by using the transfer matrix method to derive various scattering properties \cite {cantor_f1,cantor_f2,cantor_f3,cantor_f4,cantor_f5,cantor_f6,cantor_f7,cantor_f7_1,cantor_f8,cantor_f9,mh_spp}. The composition properties of the transfer matrix have been used to derive the scattering coefficients and associated properties. In Cantor fractal potential, scattering coefficients have been found to obey scaling law with the function of the wave vector $k$ \cite {cantor_f1,cantor_f2,cantor_f6, cantor_f7, cantor_f7_1}. In a recent study, we have introduced the concept of super periodic potential (SPP) \cite{mh_spp} and have derived the close form expressions of transmission and reflection coefficients for an SPP of arbitrary order provided the transfer matrix of the `unit cell' potential which is the building block of the SPP system is known. This concept of SPP has been used to derive the close form expressions for transmission amplitudes from general Cantor and standard SVC potential  \cite{mh_spp}. In the case of standard SVC  potential, starting from a length $L$ of a potential of height $V$, a portion $\frac{1}{4^{G}}$ is removed from the middle at each stage $G$ of the remaining segments of potential height $V$. In the case of general SVC potential, a fraction $\frac{1}{\rho^{G}}$ is removed from the middle at each stage $G$ instead of $\frac{1}{4^{G}}$. Here $\rho$ is a real number greater than $1$. This general SVC potential is denoted as SVC($\rho$) potential. It is to be noted that unlike the case of general Cantor potential, the case of general SVC potential doesn't preserve the same scaling of its structural elements between consecutive stages. Therefore, general SVC potential doesn't belong to the case of fractal potential. However, similar to the case of general Cantor, this system belongs to the same family of geometrical constructions which are based on division of real lines through the removal of middle segments at different stages in such a way that the system remains parity symmetric. We will show that the SVC($\rho$) potential system can be generated through the concept of SPP.

In this paper, we present the close form expression of transmission coefficient from general SVC($\rho$)  potential by using the formalism of SPP. With the help of $q$-Pochhammer symbol, the close form expression of transmission coefficient $T_{G} (k)$ is derived for this potential of arbitrary stage $G$. The transmission coefficient is found to saturate with increasing order $G$. It is found that no band like signature emerges in the tunneling of matter waves from this potential configuration. Very sharp transmission resonances are found for smaller values of wave vector $k$ which may find application for the realization of sharp transmission filters. We also studied a special case when potential height $V$ changes at each stage $G$ in a manner that preserves the total area of the potential. It is shown analytically that the reflection amplitude will fall off as $\frac{1}{k^2}$ for large $k$. 
	
We organize the paper as follows: In section \ref{svc_section} we briefly discuss about the general SVC($\rho$) potential. Section \ref{spp_svc} introduces the SPP concept and also demonstrate how a general SVC($\rho$) potential of arbitrary stage  $G$ is special case of rectangular SPP of order $G$. In section \ref{svc_calc} we calculate the close form expression of transmission coefficient from general SVC($\rho$) potential of arbitrary stage $G$ and show various results. Section\ref{scaling_rr} presents the case of scattering features when potential height of SVC($\rho$) system depends upon stage $G$.  We present results and associated discussion in section \ref{results_discussions}. We have also provided an appendix at the end for the overall calculations in detail and associated mathematical clarity for section \ref{svc_calc}.

\section{General Smith-Volterra-Cantor  potential}
\label{svc_section}
\begin{figure}
		\begin{center}
			\includegraphics[scale=0.5]{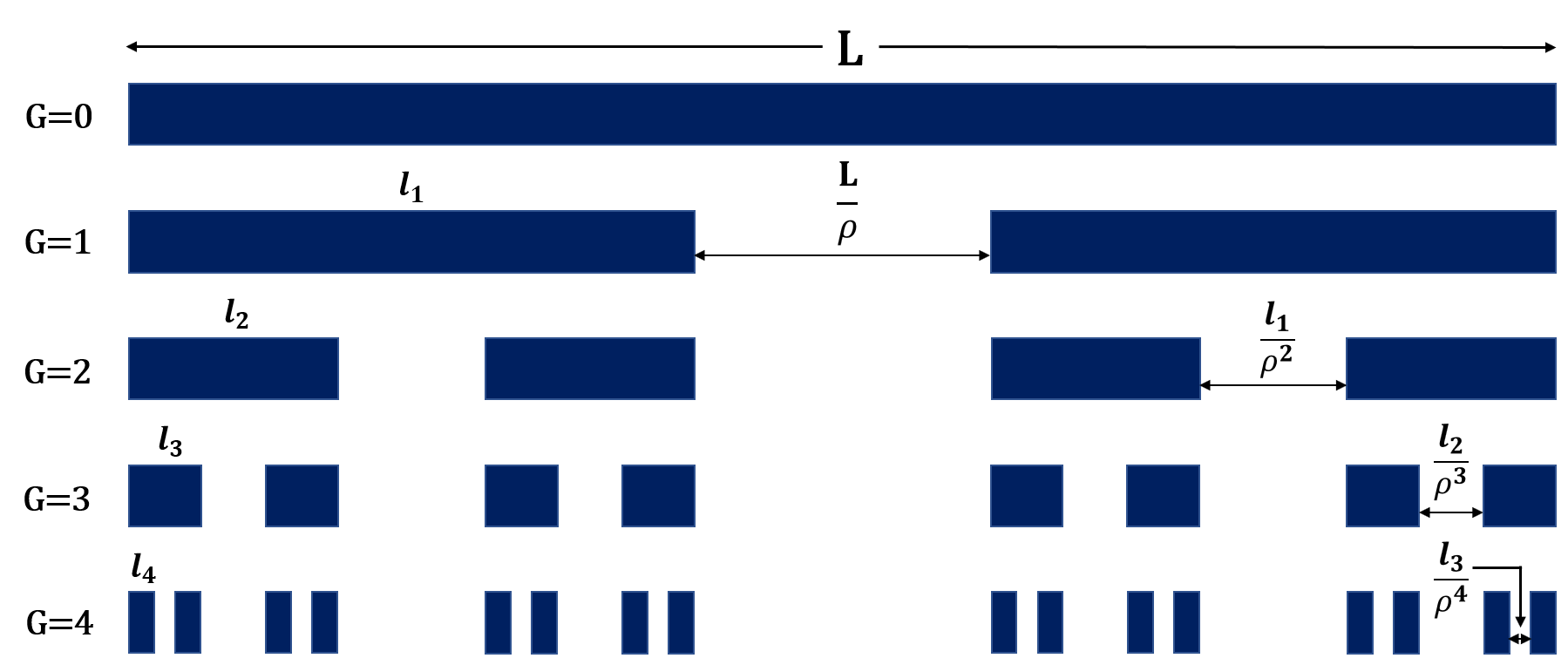} 
			\caption{\it The construction of general Smith-Volterra Cantor potential. Here the white region shows the gap between the potential and the height $V$ of blue region is the potential height. Note that consecutively lesser fractions of the previous segments are removed at a stage $G$.} 
			\label{svc_figure}
		\end{center}
	\end{figure}

	The general Smith-Volterra-Cantor system, SVC($\rho$), is a member of the family of symmetric potential. Starting from a rectangular barrier of height $V$ and length $L$, the SVC($\rho$)  potential is constructed by removing $\frac{1}{\rho^{G}}$ fraction from the middle of each remaining segments at every stage $G$. Here $\rho$ is a real number and $\rho >1$. The construction of SVC($\rho$) potential is shown in Fig-\ref{svc_figure}. At stage $G=1$, a fraction $\frac{1}{\rho}$ is removed from the middle of the length $L$ as shown in Fig-\ref{svc_figure}. In the next stage $G=2$, further a fraction $\frac{1}{\rho^{2}}$ is removed from the remaining two segments of stage $G=1$. In stage $G=3$, a fraction $\frac{1}{\rho^{3}}$ is removed from the remaining four segments of stage $G=2$. This process of removal of a fraction $\frac{1}{\rho^{G}}$ from the middle portion of the remaining segments can continue to an arbitrary stage $G$  to obtain SVC($\rho$)  potential of stage $G$. In Fig-\ref{svc_figure}, the construction of SVC($\rho$)  potential is shown upto stage $G=4$. 
	At any stage $G$, the SVC($\rho$)  potential contain $2^{G}$ segments of equal length $l_{G}$. The length $l_{G}$ can be calculated as follows: 
	
	From Fig-\ref{svc_figure}, it is clear that the segment length $l_{1}$ for $G=1$ is, 
	\begin{equation}
		l_{1} = \frac{L}{2}\left(1-\frac{1}{\rho}\right).
	\end{equation}
	Similarly, the segment length $l_{2}$ for stage $G=2$ is,
	\begin{equation}
		l_{2} = \frac{l_{1}}{2}\left(1-\frac{1}{\rho^{2}}\right) = \frac{L}{2^{2}}\left(1-\frac{1}{\rho}\right)\left(1-\frac{1}{\rho^{2}}\right).
	\end{equation}
	Similarly,
	\begin{equation}
		l_{3} = \frac{l_{2}}{2}\left(1-\frac{1}{\rho^{3}}\right) = \frac{L}{2^{3}}\left(1-\frac{1}{\rho}\right)\left(1-\frac{1}{\rho^{2}}\right)\left(1-\frac{1}{\rho^{3}}\right),
	\end{equation}
	and
	\begin{equation}
		l_{4} = \frac{l_{3}}{2}\left(1-\frac{1}{\rho^{4}}\right) = \frac{L}{2^{4}}\left(1-\frac{1}{\rho}\right)\left(1-\frac{1}{\rho^{2}}\right)\left(1-\frac{1}{\rho^{3}}\right)\left(1-\frac{1}{\rho^{4}}\right).
	\end{equation}
	By continuing the same steps, the segment length $l_{G}$ for arbitrary $G^{th}$ order SVC($\rho$)  potential is obtained as,
	\begin{equation}
		l_{G} = \frac{L}{2^{G}}\prod_{j=1}^{G}\left(1-\frac{1}{\rho^{j}}\right). 
		\label{l_G}
	\end{equation}
	The product series can be recognized as, 
\begin{equation}
		\prod_{j=1}^{G}\left(1-\frac{1}{\rho^{j}}\right) = q \left ( \frac{1}{\rho}; \frac{1}{\rho} \right)_{G}.
\end{equation}
	Where,
	\begin{equation}
		q(a;\sigma)_{n}=\prod_{j=0}^{n-1}(1-a.\sigma^{j})=(1-a)(1-a.\sigma)(1- a. \sigma^{2}).....(1-a.\sigma^{n-1})
		\label{qp}
	\end{equation} 
	is $q$-Pochhammer symbol. Through the use of $q$-Pochhammer symbol we can express $l_{G}$ as,
	
	\begin{equation}
		l_{G} = \frac{L}{2^{G}} q \left ( \frac{1}{\rho}; \frac{1}{\rho} \right)_{G} .
		\label{l_G_qp}
	\end{equation}
	In the next section we show that starting from a `unit cell' rectangular potential of height $V$ and length $l_{G}$, one can construct SVC($\rho$) potential of order $G$ as special case of rectangular SPP of order $G$.   
	\section{SVC($\rho$) potential as special case of super periodic rectangular potential}
	\label{spp_svc}
\begin{figure}[H]
		\begin{center}
			\includegraphics[scale=1.0]{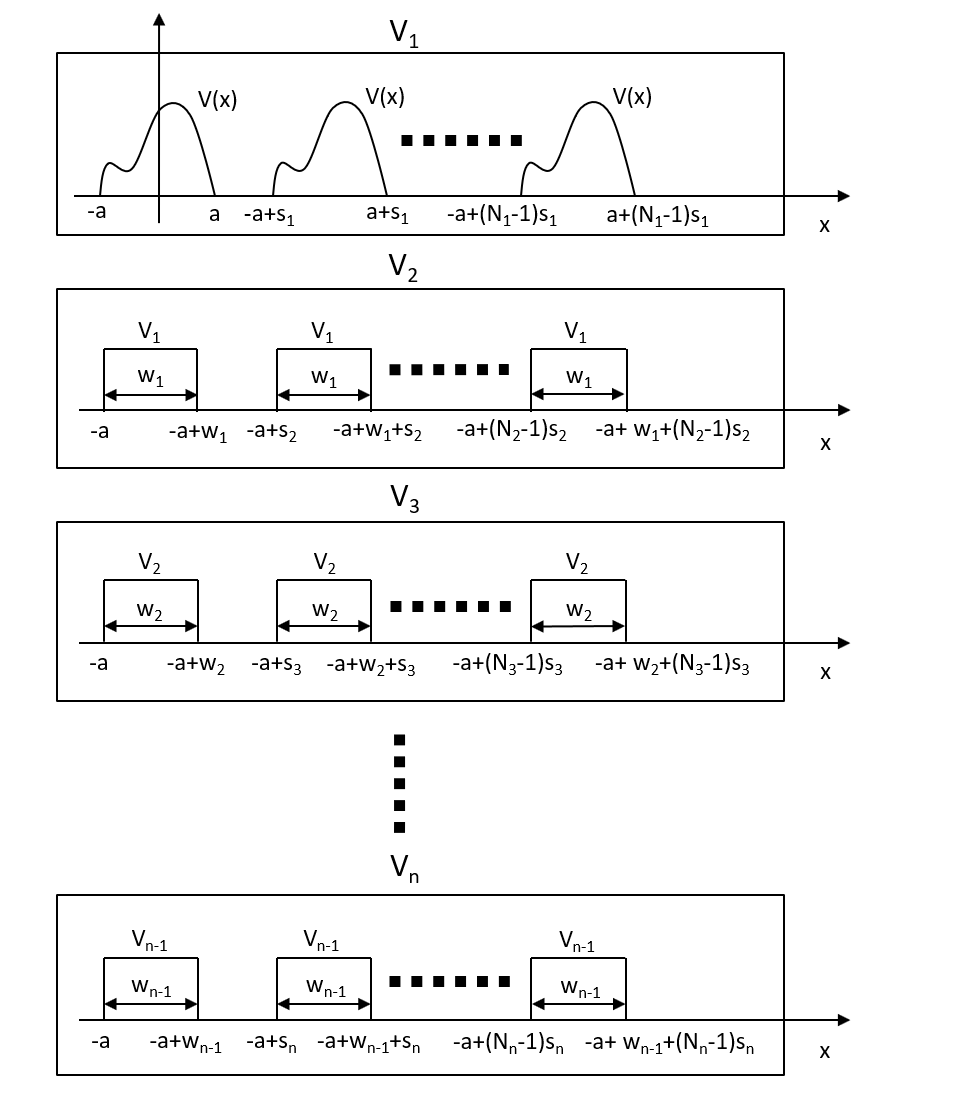}
		\end{center}
		\caption{\it Illustration of a super periodic potential of order $n$. Here $V(x)$ is the `unit cell' potential. }
		\label{superperiodicpotential}
	\end{figure}

	Super periodic potential (SPP) is the most generalized form of periodic potential. The concept of super periodic potential in quantum mechanics and its application for symmetric fractal system is introduced in \cite{mh_spp} in detail. However, for the sake of completeness of the present work, we introduce the SPP concept here. Consider a localized one dimensional  potential $V(x)$ of arbitrary form confined from $x=-a$ to $x=+a$ as shown in Fig-\ref{superperiodicpotential}. By repeating the potential $N_{1}$ times periodically at equal distances $s_{1}=2a+c_{1}$, $c_{1} \geq 0$, as shown in Fig-\ref{superperiodicpotential}, we obtain a periodic potential having `unit cell' potential as $V(x)$. Now consider the $N_{1}$ times periodically repeated collection of unit cell potential $V(x)$ as a single `unit cell' of total span $w_{1}$. This new `unit cell' potential which we can call as $V_{1}$ is further periodically repeated $N_{2}$ times with separation $s_{2}$ $\geq$ $w_{1}$ as shown in Fig-\ref{superperiodicpotential}. In the next, we consider this new potential system of total span $w_{2}$ as $V_{2}$. Now $V_{2}$ again is periodically repeated $N_{3}$ times at arbitrary chosen equal interval $s_{3}$ $\geq$ $w_{2}$ to obtain the potential $V_{3}$. This process of considering every new periodic system (a collection of unit cells at regular intervals) again as a `unit cell' and replicating it up to an arbitrary number $N_{n}$ times ($n\in I^{+}$) yields super periodic potential $V_{n}$ of order-$n$. From the Fig-\ref{superperiodicpotential}, it is clear that the total span of the unit cell potentials $V(x)$ (denote this as $V_{0}$), $V_{1}$, $V_{2}$, and $V_{3}$ are respectively given by,
	
	\begin{equation}
		w_{0} = 2a, \nonumber
	\end{equation}
	\begin{equation}
		w_{1} = w_{0} + (N_1 - 1)s_{1}, \nonumber
	\end{equation}
	\begin{equation}
		w_{2} = w_{1} + (N_2 - 1)s_{2}, \nonumber
	\end{equation}
	\begin{equation}
		w_{3} =  w_{2} + (N_3 - 1)s_{3}. \nonumber
	\end{equation}
	Similarly, the total span of SPP of any order-$n$ is 
	\begin{equation}
		w_n =  w_{n-1} + (N_{n} - 1)s_{n}. 
	\end{equation}
 %%%%%%%%%%%%%%
	\begin{figure}
		\begin{center}
			\includegraphics[scale=0.5]{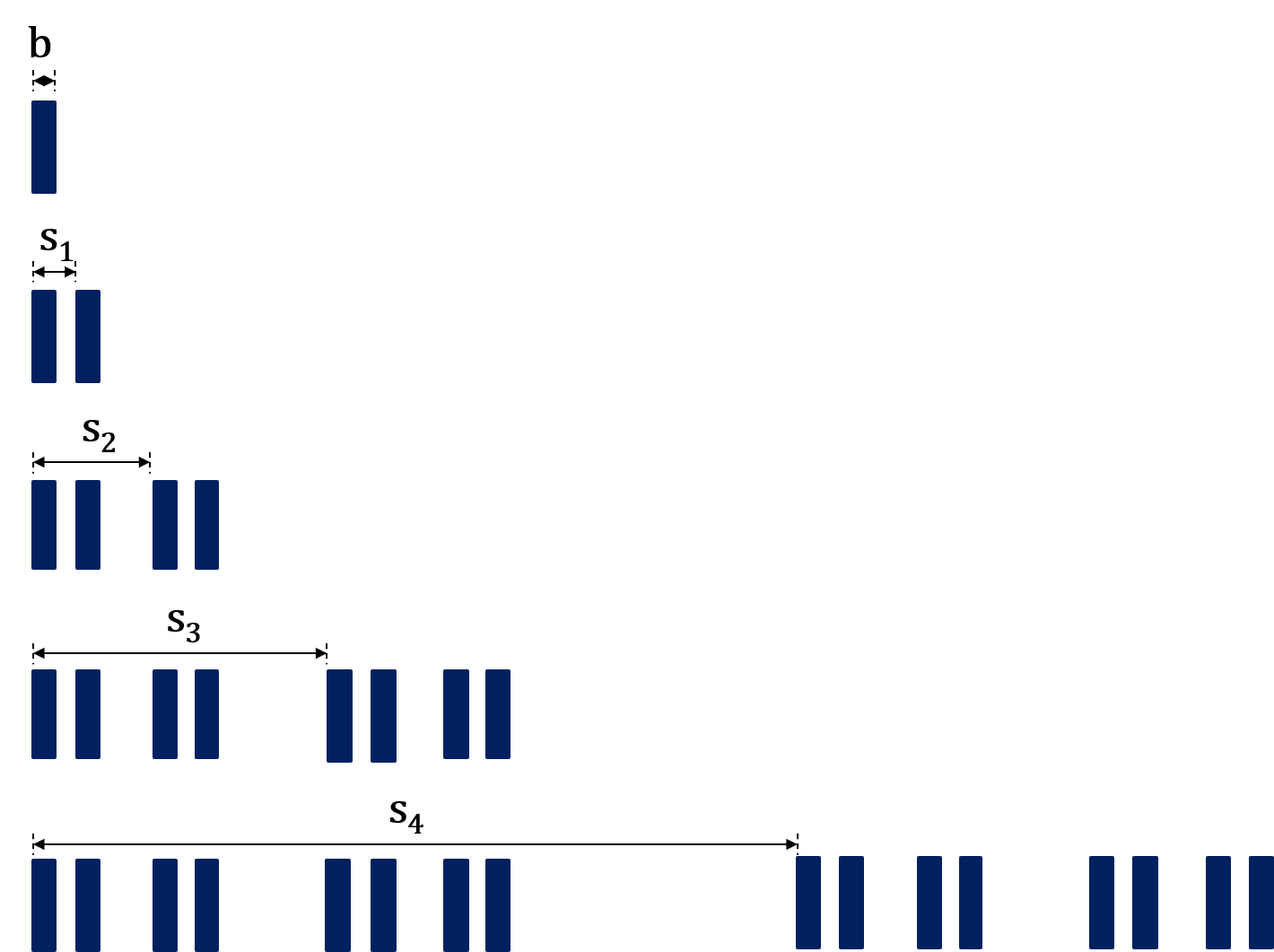} 
			\caption{\it This figure illustrate the construction of  stage $G=4$ Smith-Volterra-Cantor potential as special case of super periodic potential of order $4$ with $N_{i}=2$, $i \in I^{+}$.}
			\label{svc_c}
		\end{center}
\end{figure}
Next we show that SVC($\rho$)  potential is a special case of rectangular SPP. Here by rectangular SPP, we mean that the `unit cell' potential is a rectangular barrier potential.  Formation of SVC($\rho$)  potential up to  stage $G = 4$ is illustrated in Fig-\ref{svc_c}. Starting from a rectangular barrier of width $b=l_{G}$ with $N_{i}=2$, $ i \in$ $\lbrace$ ${1,2,3,..., G}$ $\rbrace$ and choosing suitable values of various regular intervals $s_{1}, s_{2}, s_{3},...,s_{G}$ one can construct a SVC($\rho$)  potential by performing the periodic operation on the basic `unit cell'. As can be seen from Fig-\ref{svc_c}, the construction of SVC($\rho$)  potential starts with the `initiator' as rectangular barrier and the mathematical operation performed is the super periodic operation. This is different than generating SVC($\rho$)  potential as depicted in Fig-\ref{svc_figure} in which the `initiator' is large rectangular barrier of length $L$ and the mathematical operation is suitable division of this length at different stage $G$. Further, it is easy to see that SVC($\rho$)  potential of stage $G$ has the same order $G$ of the special case of rectangular SPP.    
	
	Knowledge of $b=l_{G}$ (which is given by Eq. (\ref{l_G})) and various intervals $s_{1}, s_{2}, s_{3},...,s_{G}$ are important to construct the desired SVC($\rho$) potential. The calculations for various $s_{p}$, ($ p \in$ $\lbrace$ ${1,2,3,..., G}$ $\rbrace$) are shown below. From the definition of SVC($\rho$) potential, it is clear that,
	\begin{equation}
		s_{1} =l_{G}+\frac{l_{G-1}}{\rho^{G}}.  \nonumber
	\end{equation}
	Similarly,
	\begin{equation}
		s_{2} = l_{G-1}+\frac{l_{G-2}}{\rho^{G-1}},   \nonumber
	\end{equation}
	\begin{equation}
		s_{3} = l_{G-2}+\frac{l_{G-3}}{\rho^{G-2}}.   \nonumber
	\end{equation}
	Therefore for a general $s_{p}$ we have the following expression,
	\begin{equation}
		s_{p} =  l_{G+1-p}+\frac{l_{G-p}}{\rho^{G+1- p}}.
	\end{equation}
	Using Eq. (\ref{l_G}) in above expression, we have
	\begin{equation}
		s_{p}=\frac{L}{2^{G+1-p}}\prod_{j=1}^{G+1-p}\left(1-\frac{1}{\rho^{j}}\right)+\frac{L}{\rho(2\rho)^{G-p}}\prod_{j=1}^{G-p}\left(1-\frac{1}{\rho^{j}}\right).
	\end{equation}
	\begin{equation}
		s_{p}=\frac{L}{2^{G+1-p}}\left(1-\frac{1}{\rho^{G+1-p}}\right)\prod_{j=1}^{G-p}\left(1-\frac{1}{\rho^{j}}\right)+\frac{2L}{(2\rho)^{G+1-p}}\prod_{j=1}^{G-p}\left(1-\frac{1}{\rho^{j}}\right).
	\end{equation}
	Further simplifications leads to,
	\begin{equation}
		s_{p}=\frac{L}{2^{G+1-p}}\left(1+\frac{1}{\rho^{G+1-p}}\right)\prod_{j=1}^{G-p}\left(1-\frac{1}{\rho^{j}}\right).
		\label{s_i}
	\end{equation}
	In Eq. (\ref{s_i}) the product term can be written in the form of $q$-Pochhammer symbol as defined earlier in Eq. (\ref{qp}). Therefore,
	\begin{equation}
		s_{p}=\frac{L}{2^{G+1-p}}\left(1+\frac{1}{\rho^{G+1-p}}\right)\times q \left (\frac{1}{\rho}; \frac{1}{\rho} \right )_{G-p}.
		\label{s_i_qp}
	\end{equation}
	\section{Tunneling probability from SVC($\rho$)  potential}
	\label{svc_calc}
	Reference \cite{mh_spp} has shown in detail that if the transfer matrix of `unit cell' potential is known then one can obtain the transfer matrix of the corresponding super periodic potential (SPP) of any arbitrary order-$n$. From the knowledge of the transfer matrix, the reflection and transmission coefficients are easily obtained. If the transfer matrix $M$,
	\begin{equation}
		M (k)= \begin{pmatrix}   M_{11} (k) & M_{12} (k) \\ M_{21} (k) & M_{22} (k)   \end{pmatrix}  
	\end{equation}
	of a one dimensional finite range potential is known, then the transmission amplitudes for the corresponding SPP of order-$n$ is given by \cite{mh_spp}, 
	\begin{equation}
		T(N_{1},N_{2},...,N_{n})=\frac{1}{1+[|M_{12}|U_{N_{1}-1}(\zeta_{1})U_{N_{2}-1}(\zeta_{2})U_{N_{3}-1}(\zeta_{3})........U_{N_{n}-1}(\zeta_{n})]^{2}}. 
		\label{t_spp}
	\end{equation}
	In the above equation,  $U_{N}(\zeta)$ is the Chebychev polynomial of the second kind \cite{abramowitz1964} and various $\zeta_{i}$s  appearing in the above equation are the Bloch phase of the corresponding fully developed periodic system. Starting from the transfer matrix elements, the calculation of various $\zeta_{i}$s are shown in detail in the Appendix-A at the end.  The elements of the transfer matrix for a rectangular barrier of width $b$ which is located in interval $(-\frac{b}{2},\frac{b}{2})$  \cite{griffith},
	\begin{subequations}
		\begin{equation}
			M_{11}=(\cos{\kappa b}-i\varepsilon_{+} \sin{\kappa b})e^{ikb},
		\end{equation}
		\begin{equation}
			M_{12}=i\varepsilon_{-} \sin{\kappa b},
		\end{equation}
		\begin{equation}
			M_{21}=-i\varepsilon_{-} \sin{\kappa b},
		\end{equation}
		\begin{equation}
			M_{22}=(\cos{\kappa b}+i\varepsilon_{+} \sin{\kappa b})e^{-ikb}.
		\end{equation}
	\end{subequations}
	
Where,
    \begin{equation}
		\varepsilon_{\pm}=\frac{1}{2} \Big( \mu\pm \frac{1}{\mu}\Big),
		\label{epsilon_plus_minus}
		\nonumber
	\end{equation}
	\begin{equation}
		\mu=\frac{k}{\kappa} \ \ , \ \ k=\frac{\sqrt{2mE}}{\hbar},  \nonumber 
	\end{equation} 
	and,
	\begin{equation}
		\kappa=\frac{\sqrt{2m(E-V)}}{\hbar}. \nonumber
	\end{equation}
	For a known transfer matrix, the expression for $\zeta_{1}$ and $\zeta_{2}$ are given by,
	\begin{equation}
		\zeta_{1}= \vert M_{22}\vert \cos{(\theta + ks_{1})},
	\end{equation}
	\begin{equation}
		\zeta_{2}=\lvert{M_{22}}\rvert U_{N_{1}-1}(\zeta_{1})\cos{\big[\theta-k\{(N_{1}-1)s_{1}-s_{2}\}\big]}-U_{N_{1}-2}(\zeta_{1})\cos{\big[k(N_{1}s_{1}-s_{2})\big]},
		\label{zeta2}
	\end{equation}
	where $\theta$ is the argument of $M_{22}$. Various other $\zeta_{j}$s for $ j \in \{3,4,..,G \}$ can be calculated from the following expression, 
	\begin{multline}
		\zeta_{j}(k) = \vert M_{22} \vert \cos \left[\theta - k \left \{ \sum_{p=1}^{j-1}(N_{p}-1)s_{p} - s_{j}\right \} \right ] {\prod_{p=1}^{j-1}U_{N_{p}-1}(\zeta_{p})} \\ -\sum_{r=1}^{j-2}\left [\cos \left\{k\left({\sum_{p=r}^{j-1}N_{p}s_{p}} - {\sum_{p=r+1}^{j}s_{p}}\right)\right\}U_{N_{r}-2}(\zeta_{r}){\prod_{p=r+1}^{j-1}U_{N_{p}-1}(\zeta_{p})}\right ]\\-U_{N_{j-1}-2}(\zeta_{j-1})\cos[k(N_{j-1}s_{j-1}-s_{j})].
		\label{zeta_nn}
	\end{multline}
%%%%%%%%%%%%%%%%%%%%%%%%%%%%%%%%%%%%%%%%%%%%%%%%%%	
	As discussed in the previous section, SVC($\rho$)  potential of stage $G$ is a super periodic rectangular potential of order $n=G$ with $N_{j}=2$, $j \in \{ 1,2,3,...,G\}$ where the thickness of `unit cell' rectangular barrier is $b=l_{G}$ given by Eq. (\ref{l_G}). Performing simplification of Eq. (\ref{zeta_nn}) with $\forall$ $N_{j}=2$ and making use of $U_{0} (y)=1$ and $U_{1} (y)=2y$, we have
\begin{equation}
		\zeta_{j}(k) = 2^{j-1}\vert M_{22} \vert\cos\big[\alpha-k\eta_{1}(j)\big]\prod_{p=1}^{j-1} \zeta_{p} - \sum_{r=1}^{j-1}\left[2^{j-r-1}\cos \big[k\eta_{2}(j,r)\big] \prod_{p=r+1}^{j-1}\zeta_{p}\right].
		\label{zeta_n1}
\end{equation}
	In the above, 
	\begin{equation}
		\eta_{1}(j) = \left(\sum_{p = 1}^{j-1}s_{p}\right)-s_{j},
	\end{equation}
	and  $\eta_{2}(j,r)$ is given by,
	\begin{equation}
		\eta_{2}(j,r)  =\left(\sum_{p = r}^{j}s_{p}\right)- ( 2s_{j}-s_{r}).
	\end{equation}
%%%%%%%%%%%%%%%%%%%%%%%%%%%%%%%%%%%%%%%%%%%%%%%%%%
	Now with the knowledge of $\zeta_{1}$, $\zeta_{2}$, $\zeta_{3}$,.....,$\zeta_{G}$ we can compute the transmission coefficient by using Eq. (\ref{t_spp}). Substituting $N_{j}=2$ in Eq. (\ref{t_spp}), we get for SVC($\rho$) potential
	\begin{equation}
		T_{G}(k)=\frac{1}{1+\varepsilon_{-}^{2}\sin^{2}{(\kappa l_{G})} [U_{1}(\zeta_{1})U_{1}(\zeta_{2})U_{1}(\zeta_{3})....U_{1}(\zeta_{G})]^{2} }.
	\end{equation} 
	Using $U_{0}(y)=1$ and $U_{1}(y)=2y$ we simplify the above expression to obtain,
	\begin{equation}
		T_{G}(k)=\frac{1}{1+4^{G}\varepsilon_{-}^{2}\sin^{2}{(\kappa l_{G})} \prod_{j=1} ^{G} \zeta_{j}^{2}},
		\label{T_svc_rho}
	\end{equation}
	where $l_{G}$ is given by Eq. (\ref{l_G}) and $\zeta_{j}$s from Eq. (\ref{zeta_n1}).
\begin{figure}[H]
		\begin{center}
			\includegraphics[scale=0.20]{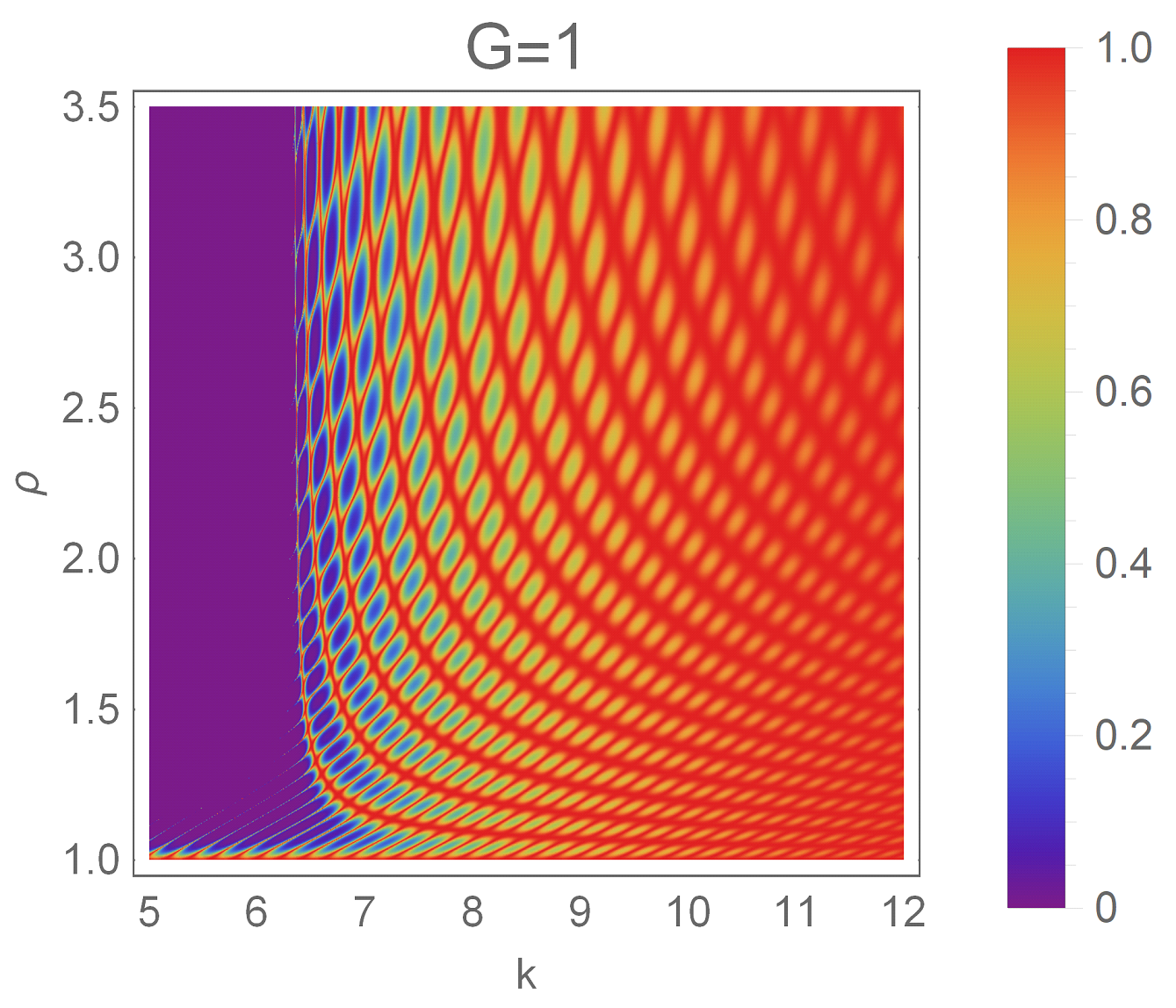} (a) \ \includegraphics[scale=0.20]{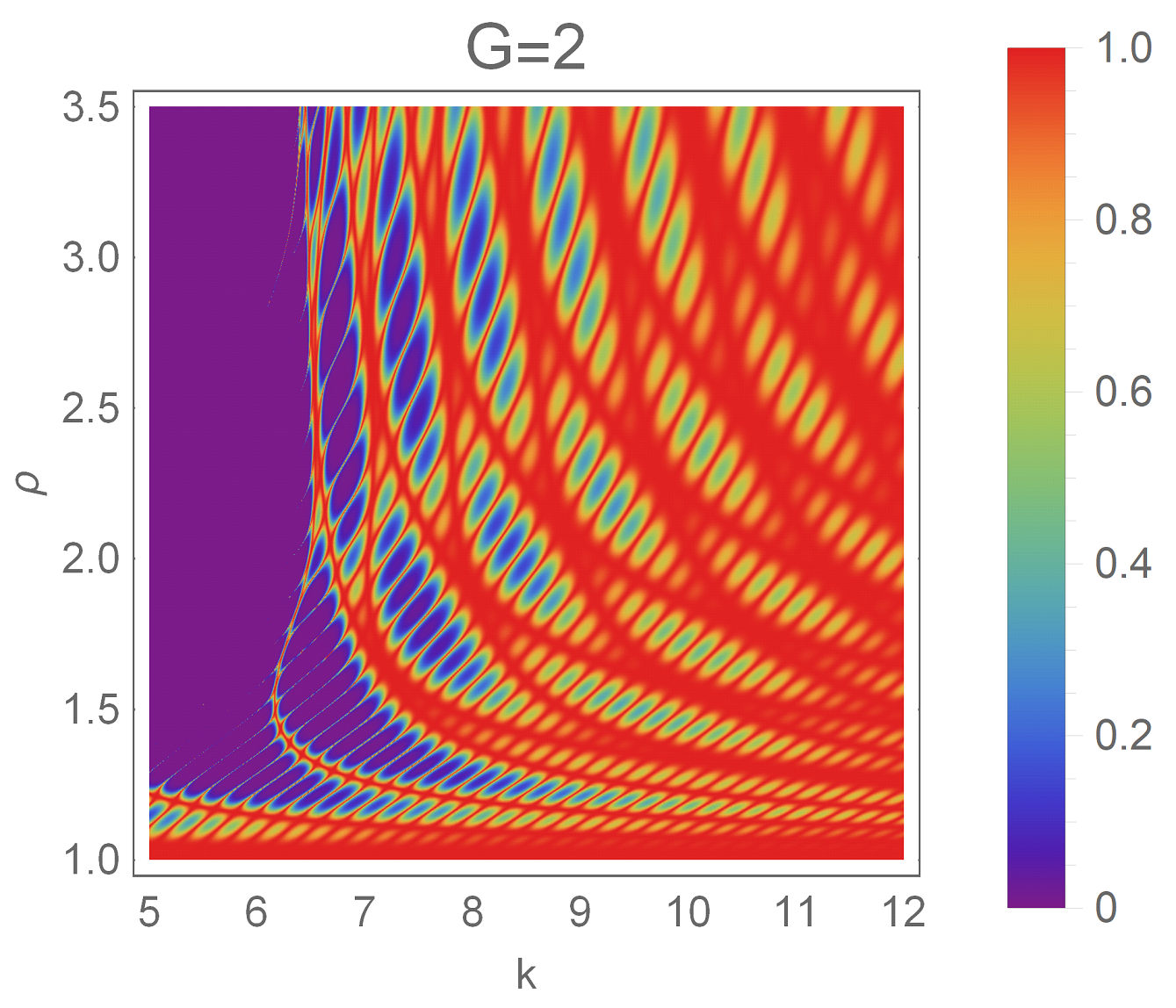}  (b) \\
			\includegraphics[scale=0.20]{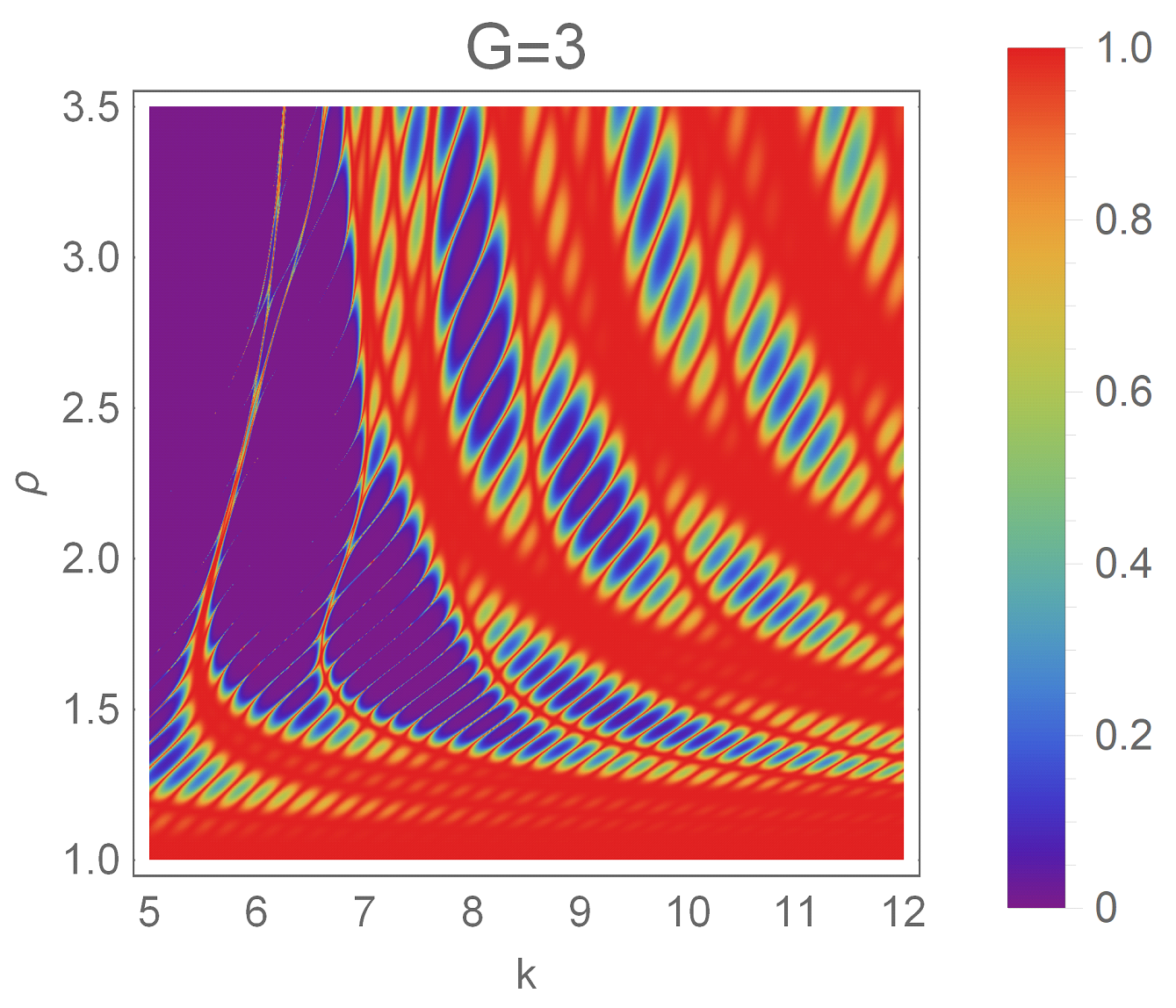} (c) \ \includegraphics[scale=0.20]{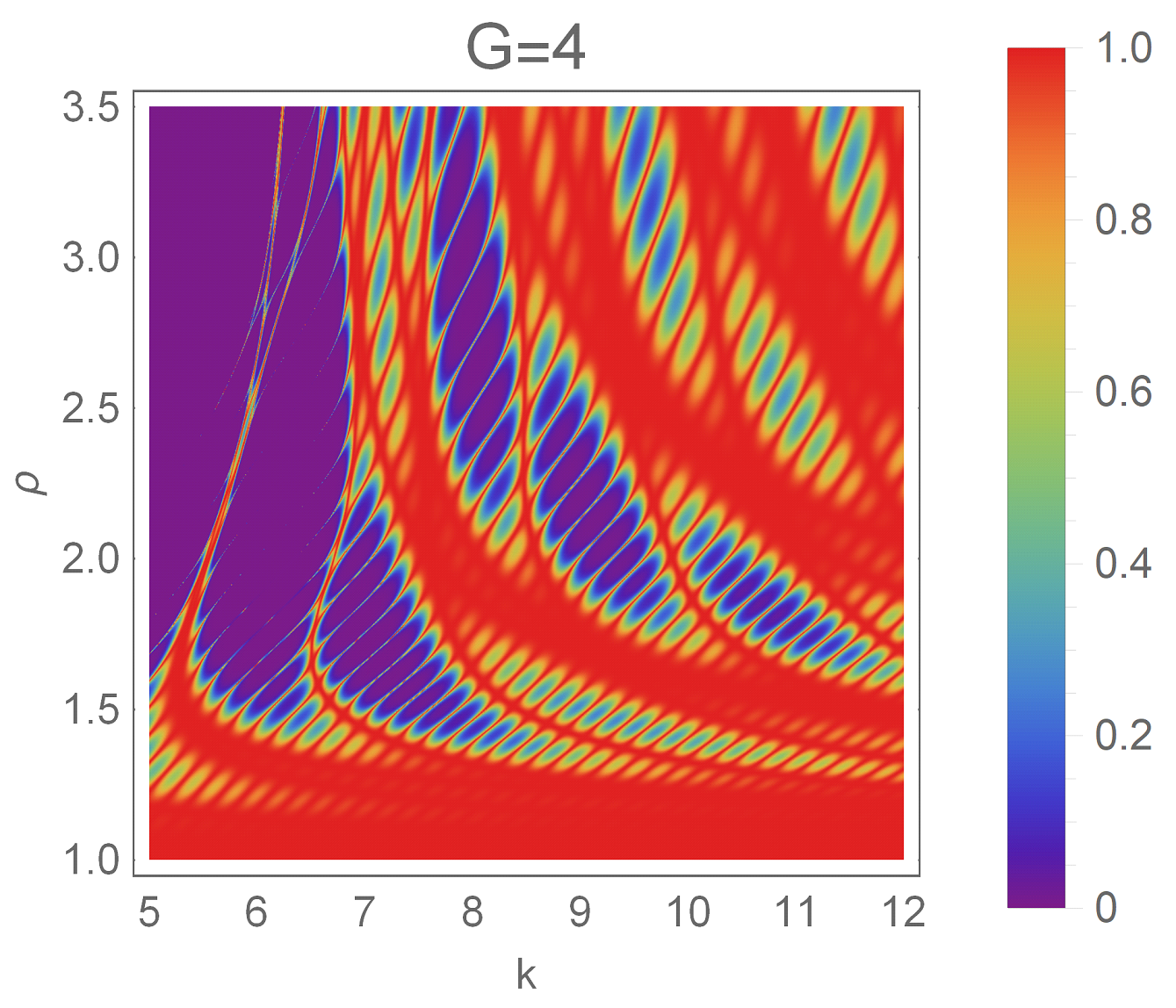}  (d) \\
			\includegraphics[scale=0.20]{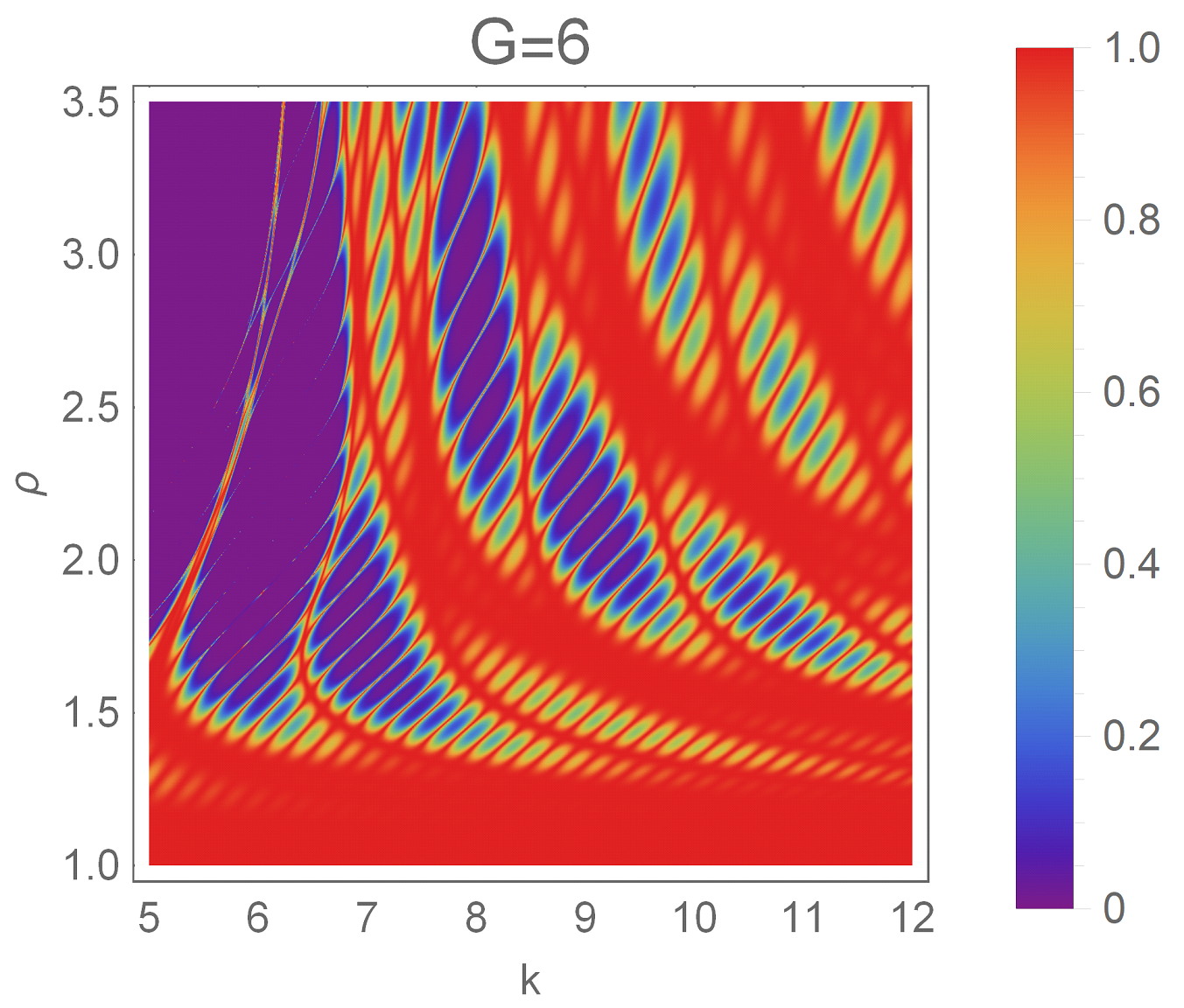} (e) \ \includegraphics[scale=0.20]{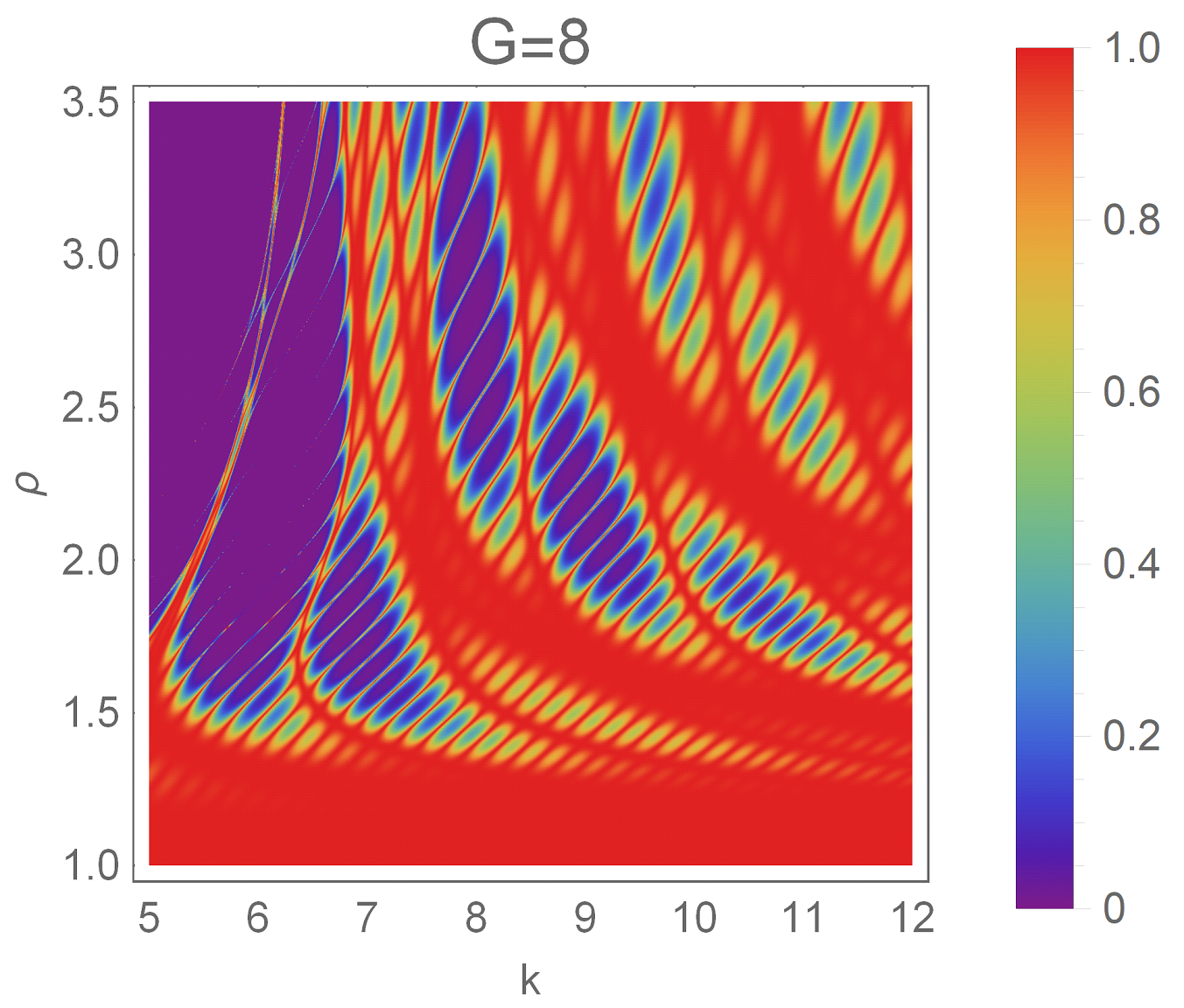}  (f)
            \end{center}
			\caption{\it Density plot of tunneling coefficient $T$ for different stage $G$ of the general SVC($\rho$)  potential of height $V=20$ and total span $L=15$. The plots shows sharp features of transmission coefficient for lower $k$ values.}  
			\label{density_plots_svc}
	\end{figure}  
	
	\begin{figure}[H]
		\begin{center}
			\includegraphics[scale=0.35]{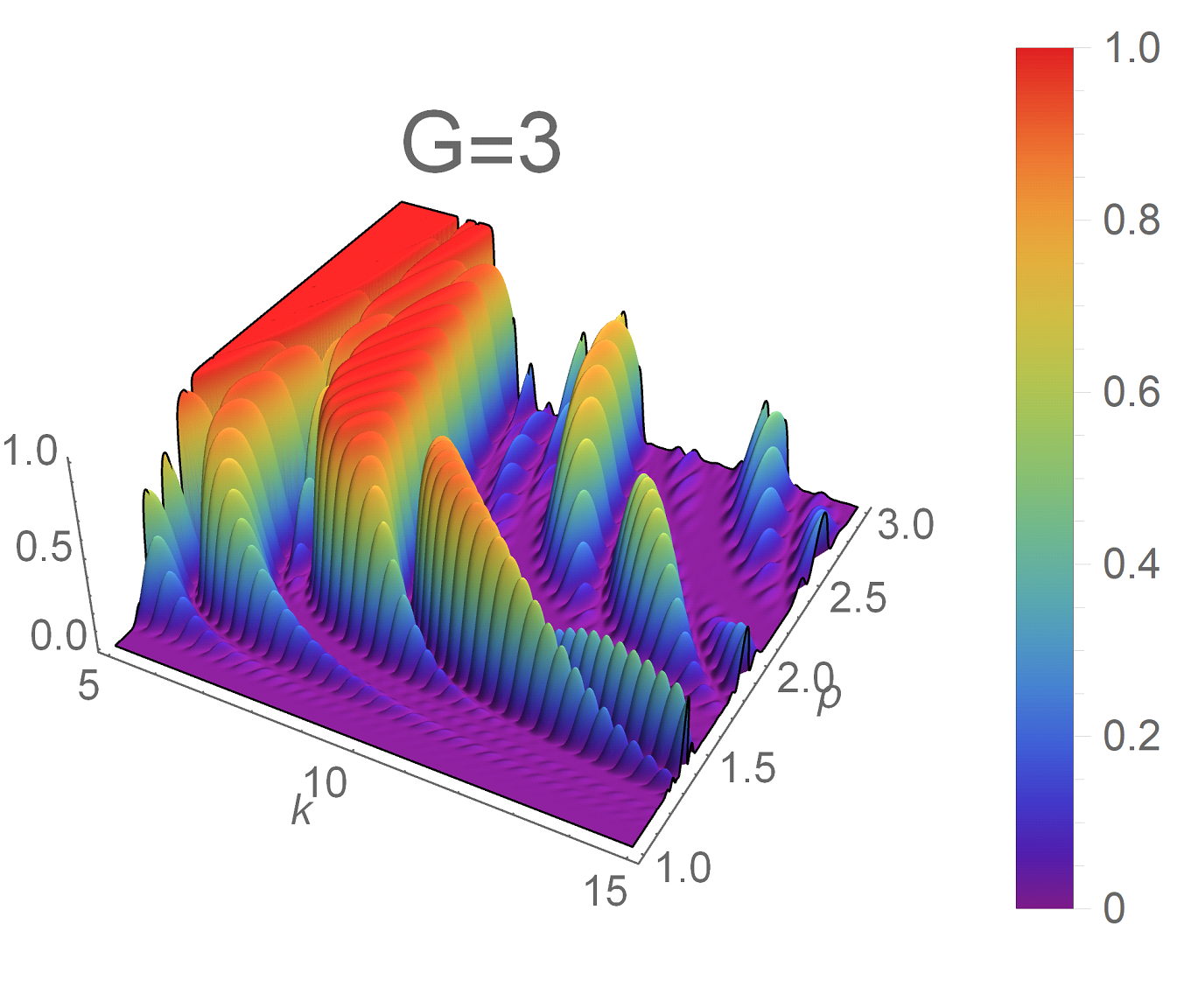} (a) \\
			\includegraphics[scale=0.35]{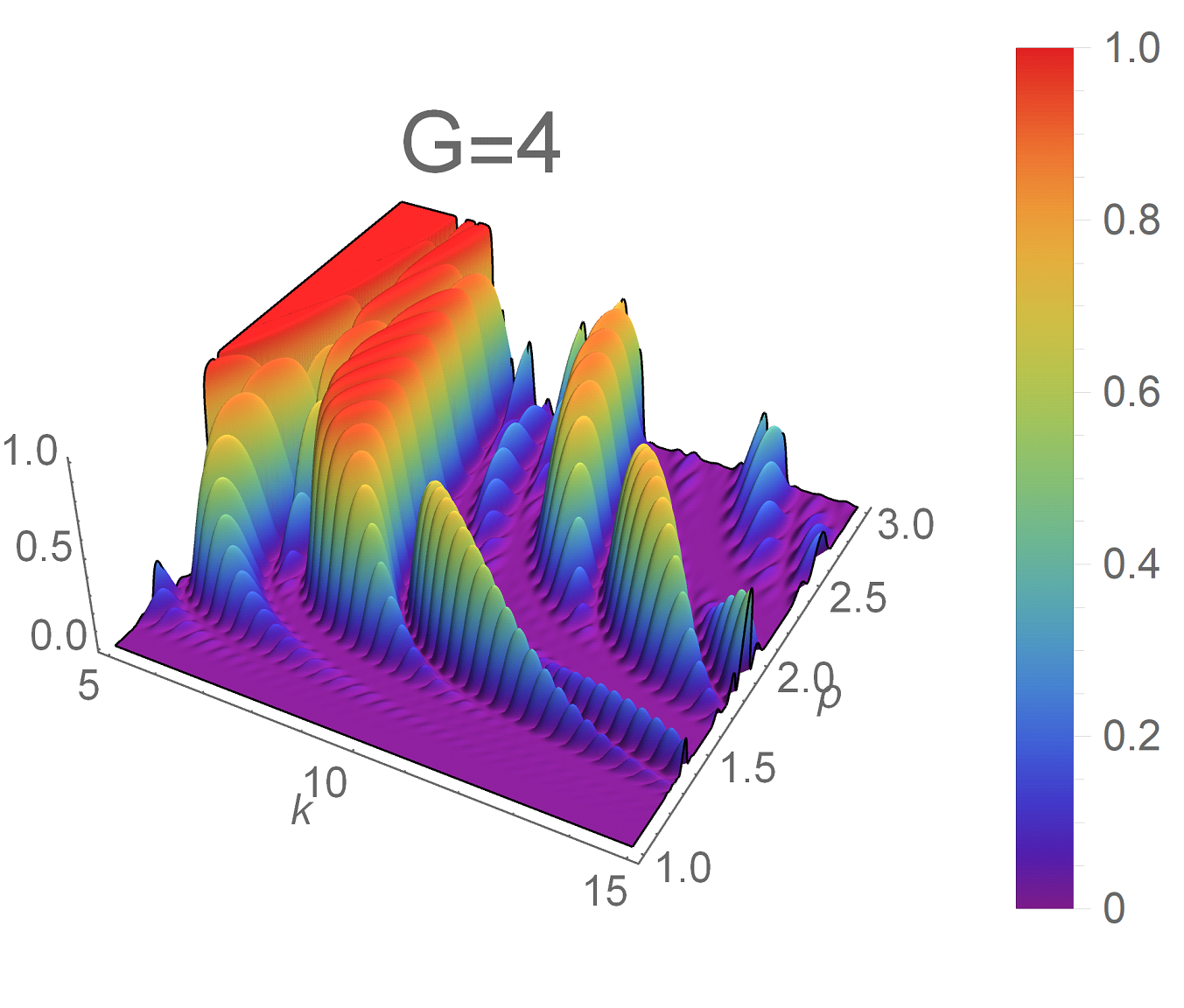}  (b) 
			\caption{\it Plots showing the reflection coefficient $R$ for stage $G=3$ and $4$ of the general SVC($\rho$)  potential with $\rho$ and $k$. Here height $V=20$ and total span $L=10$. In the $\rho-k$ plane, valleys appear for which reflection is minimal.}  
			\label{3d_plots_svc}
		\end{center}
	\end{figure}  
	
	\begin{figure}[H]
		\begin{center}
			\includegraphics[scale=0.35]{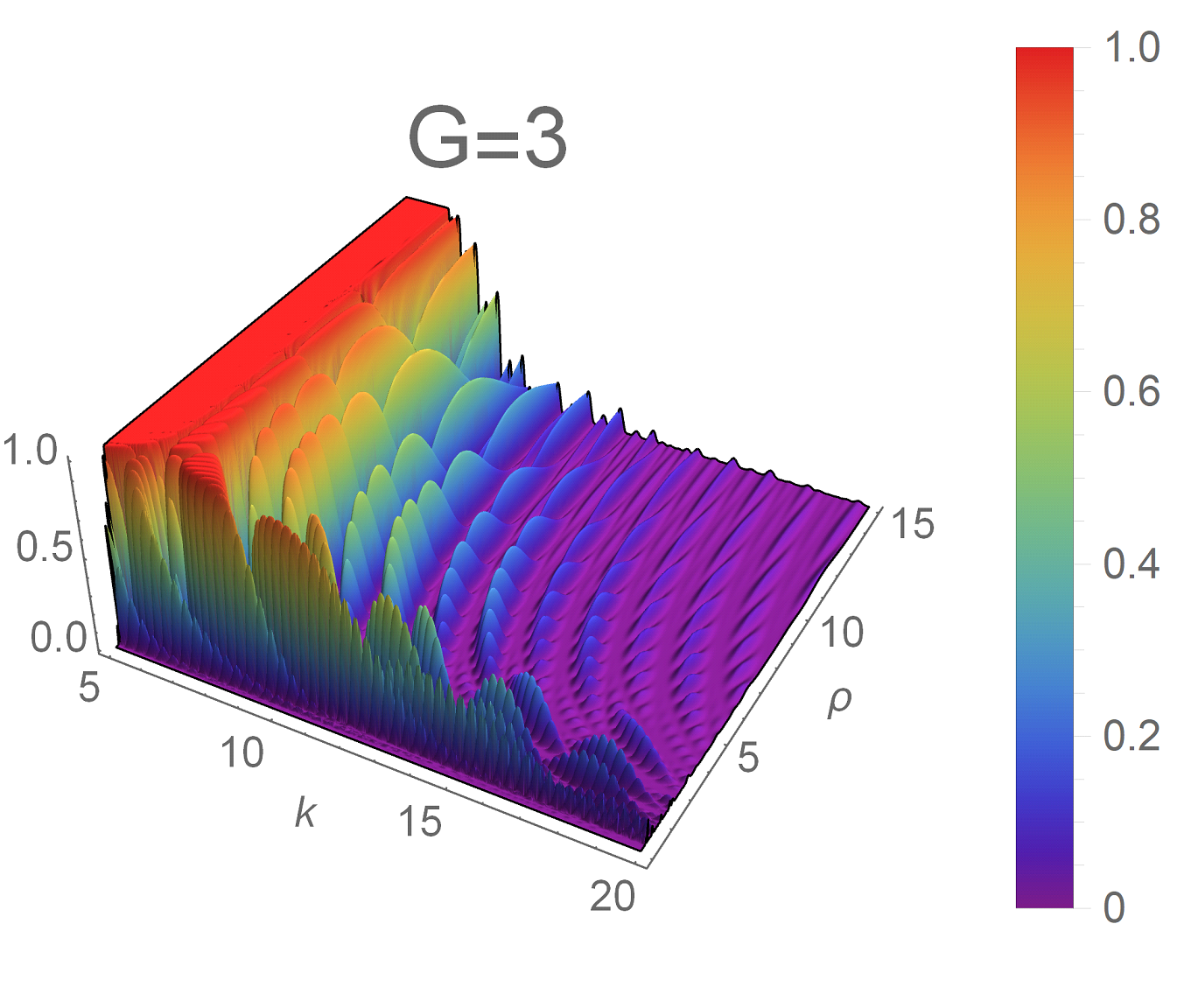} 
			\caption{\it Plot showing the reflection coefficient $R$ for stage $G=3$ and larger range of $\rho$ as compared to previous figure. Here fix parameters (V and L) are same as above Fig. (5). In the $\rho-k$ plane, valleys appear for which reflection is minimal. }  
			\label{3d_plots_svc_large_rho}
		\end{center}
	\end{figure} 
%%%%%%%%%%%%%%%%%%%%%%%%%%%%%%%%%%%%%%%%%%%%%%%%%%%%%%%%%%%%%%%%%%%%%%%%%%%%%%%%%%%%
	
	The density plot of tunneling coefficient $T_{G}(k)$ for different stage $G$ of the SVC($\rho$)  potential is shown in Fig-\ref{density_plots_svc}. From the figure it is noted that for $\rho \sim 1$, the tunneling coefficient are near to unity for all wave number $k$. This is because when $\rho$ is closer to unity, larger portion from the total span $L$ is removed. At $\rho=1$, there is no barrier at all. Also as $G$ increases, more portion from the SVC($\rho$) potential is removed and we observed more transparency for little higher value of $\rho$ beyond $1$. It is further noted in Fig-\ref{density_plots_svc} that the plots for transmission coefficients are nearly same in appearance for $G=6$ and $G=8$. The reason for this is due to the fact that consecutively lesser fraction of the previous segments are removed at each stage $G$, therefore the transmission coefficient would show a saturation with increasing stage of SVC($\rho$)  potential. This will be more evident for higher $\rho$ values. 
	
	Fig-\ref{3d_plots_svc} shows the 3D plots of the variation of reflection coefficient $R_{G} (k)$ over $\rho -k$ plane. It is seen that in the $\rho -k$ plane, the plots show the appearance of valleys in which reflection is minimal. This is also seen in the density plots of $T_{G}(k)$ in Fig-\ref{density_plots_svc} where the large regions appears for which transmission is closer to unity. In both the plots (Fig-\ref{density_plots_svc} and \ref{3d_plots_svc}) we do not observe a sequential range of $k$ over which $T_{G}(k)$ is much closer to unity (the range is interrupted by small wavy features). Therefore, energy band like signature doesn't appear for SVC$(\rho)$ potential. We have also checked for higher $G$ values and do not observe band like features from this potential. Fig-\ref{3d_plots_svc_large_rho} shows $3$D plot of $R_{G} (k)$ for range of $\rho$ up to $15$ for $G=3$. For larger $\rho$ and $k$ there is overall reduction in $R$ with smaller wave like features. We are unable to notice any self-similarity and associated scaling features in $R_{G} (k)$ with $k$ which has been noted to occur for Cantor fractal potentials \cite{cantor_f1,cantor_f2,cantor_f6,cantor_f7}. At presently we have not performed any analytical investigation to search for the scaling relation in scattering amplitude from this general SVC$(\rho)$ system as detailed in \cite{cantor_f7} for Cantor fractal potential. The analytical limitations which we have faced for this is the lack of close form analytical expression for $\sum_{i} ^{j} q(a;\sigma)_{i}$ which is encountered in our investigations. This series doesn't has known mathematical expression to the best of our knowledge. At this stage we have also refrained ourselves to search for the self-similarity in the scattering amplitude through purely numerical investigations as the outcome may be inconclusive. However a detailed investigation for the same may be performed in future work.   
 \begin{figure}[H]
		\begin{center}
			\includegraphics[scale=0.38]{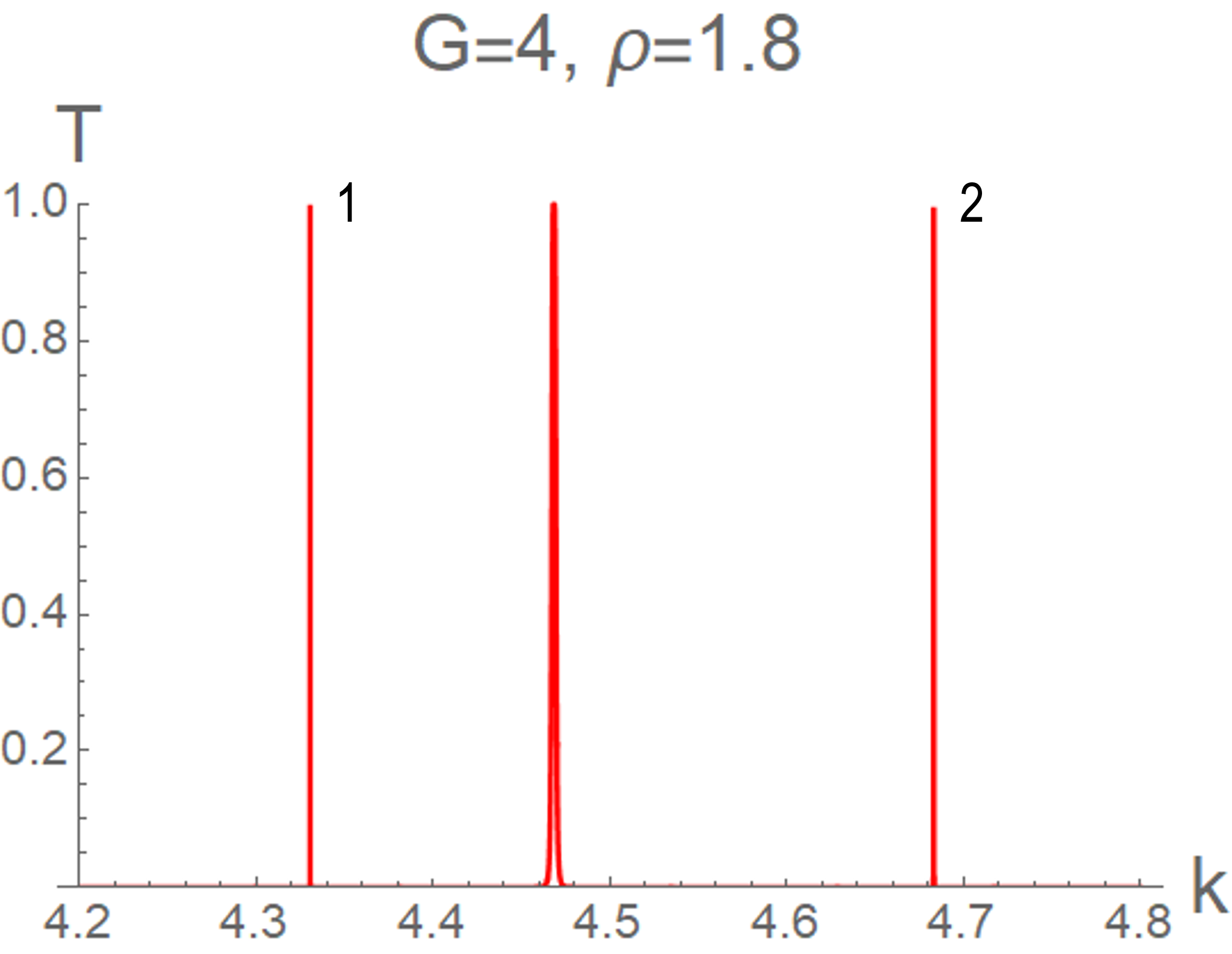} (a) 
            \includegraphics[scale=0.38]{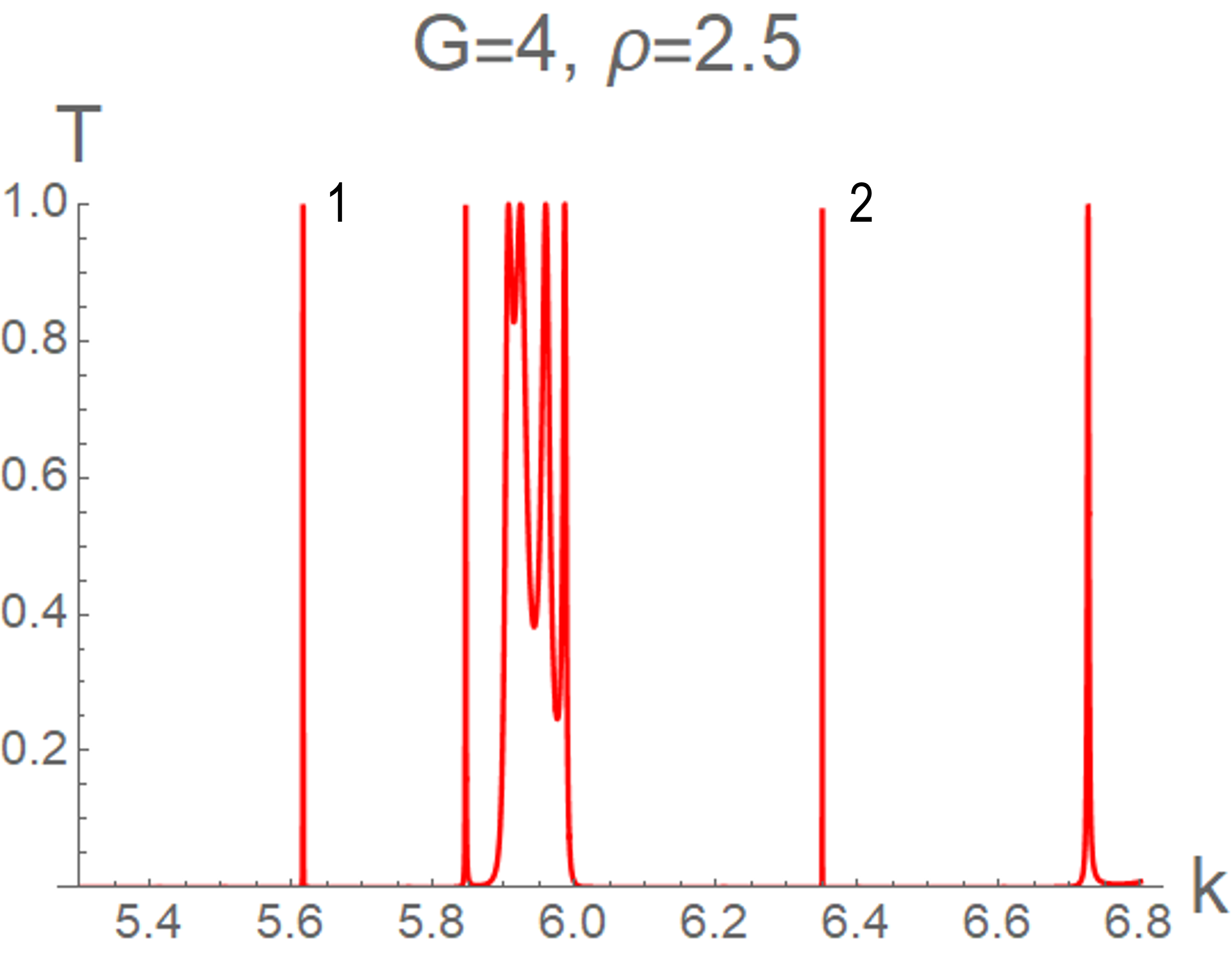} (b)
			\caption{\it Plots of the tunneling amplitudes shows the sharpness of the transmission resonances from SVC($\rho$)  potential of stage $G=4$.
			Here potential of height $V=20$ and total span $L=15$. Very sharp transmissions peaks are indicated by number $1$ and $2$ in the above figure.} 
			\label{2d_plots_sharppeaks}
		\end{center}
	\end{figure}

	An important features in tunneling coefficient from this system is the appearance of very sharp transmission resonance peaks at lower values of $k$. This is demonstrated graphically in Fig-\ref{2d_plots_sharppeaks} for different $\rho$ values for $G=4$ stage. These features can also be seen in the density plot shown in Fig-\ref{density_plots_svc} as very thin streak of lines. Due to sharpness of the transmission resonance, the density plots shows broken lines due to technical limitations of the density plots. However these are clearly evident in the 2D plots shown in Fig-\ref{2d_plots_sharppeaks}. An optical analogue of this  system may find applications in the design of sharp transmission filters.        

\section{Scaling behavior of reflection coefficient}
\label{scaling_rr}
In this section we present the scaling behavior of the reflection coefficient $R= \vert r \vert ^{2}$ with $k$ for the general SVC system. In this section we also present how the reflection coefficient behaves when height of the potential $V$ varies in a specific manner at each stage $G$. For larger $k$, reflection coefficient $R$ is very small. In this limit, $R$ can be approximated using Eq. (\ref{T_svc_rho}) as
\begin{equation}
    R \sim 4^{G}\varepsilon_{-}^{2}\sin^{2}{(\kappa l_{G})} \prod_{j=1} ^{G} \zeta_{j}^{2}.
    \label{r_small_value}
\end{equation}
Again for larger $k$ we have $\frac{V}{k^2} < < 1$ and simplification of the above expression in this limit utilizing Eq. (\ref{epsilon_plus_minus}) further leads to,
\begin{equation}
    R \sim 4^{G} \left ( \frac{1}{2} V l_{G} \right )^{2} \frac{1}{k ^{2}} \prod_{j=1}^{G} \zeta_{j}^{2}.
    \label{r_small_value}
\end{equation}
If $V_{G}$ is the height of the potential at each stage $G$, then it can be shown that the following value of $V_{G}$ keeps the total area of potential barrier (sum of the area of all potential segment at stage $G$) as constant
\begin{equation}
    V_{G} =\frac{L}{2^{G} l_{G}} V_{0},
    \label{vg_values}
\end{equation}
where $V_{0}$ is the height of the potential barrier at $G=0$ and $l_{G}$ is the width of `unit cell' rectangular barrier defined already (Eq. (\ref{l_G})). If $R_{G}$ is the reflection coefficient at each stage $G$ with the potential height of each segment as $V_{G}$ then, it can be shown that (valid for large $k$)
\begin{equation}
    \frac{R_{G}}{L^{2} V_{0}^{2}} \sim  \frac{1}{4 k ^{2}} \prod_{j=1}^{G} \zeta_{j}^{2}.
    \label{vg_zetag}
\end{equation}
\begin{figure}[H]
\begin{center}
\includegraphics[scale=0.98]{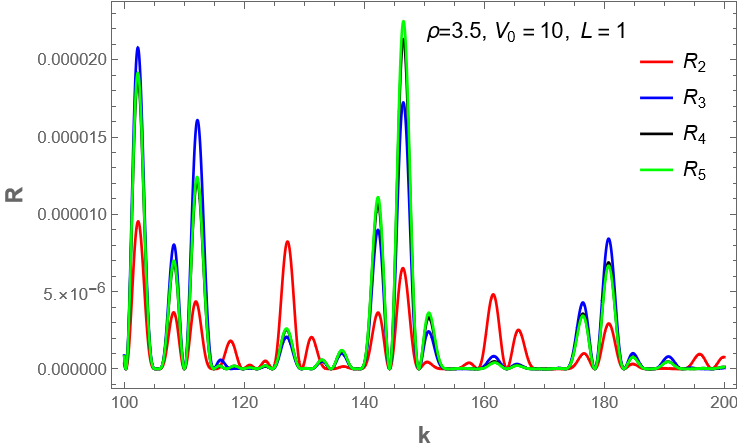} 
\caption{\textit{Plot showing the reflection coefficient for general SVC potential for $G=2$, $3$, $4$ and $5$. The potential height $V_{G}$ is determined from Eq. \ref{vg_values}. Other potential parameters are shown in the figure. It is noted that the reflection coefficient rapidly converge as $G$ increase. In other words this shows the convergence of the product term of Eq. \ref{vg_zetag} for general SVC potential.}}
\label{rg_gsvc_convergence}
\end{center}
\end{figure}

\begin{figure}[H]
\begin{center}
\includegraphics[scale=0.56]{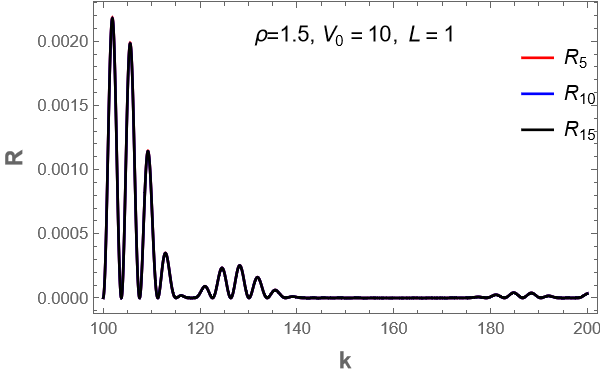} (a) \  \includegraphics[scale=0.56]{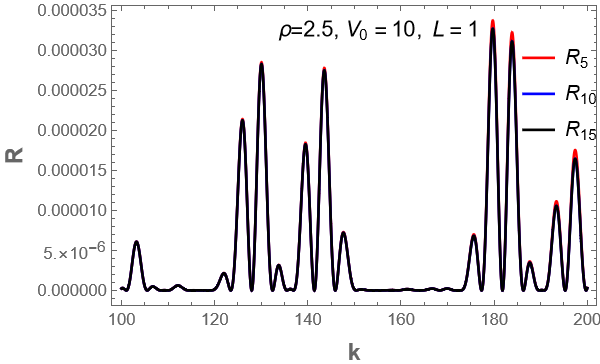} (b) \\
\includegraphics[scale=0.56]{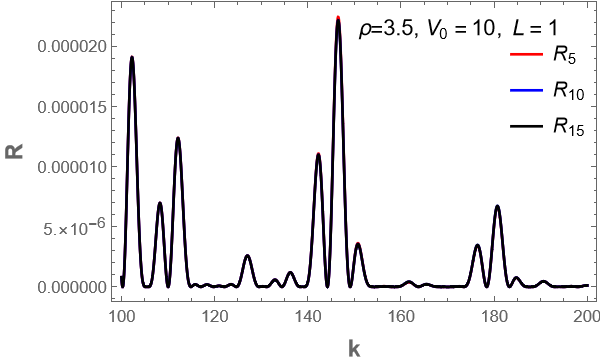} (c) \  \includegraphics[scale=0.56]{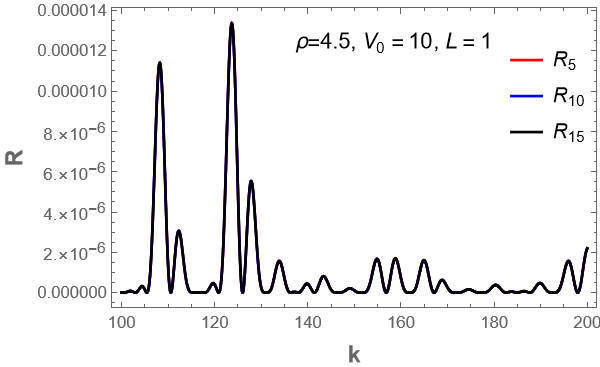} (d) \\
\caption{\textit{Plots showing the reflection coefficient for general SVC potential for $G=5, 10$ and $15$. The potential height $V_{G}$ is determined from Eq. \ref{vg_values}. Other potential parameters are shown in the figure. The difference between these three plots  is invisible. This shows the convergence of the product term of Eq. \ref{vg_zetag} for general SVC potential. }}
\label{rg_gsvc_saturation}
\end{center}
\end{figure}
\begin{figure}[H]
\begin{center}
\includegraphics[scale=0.56]{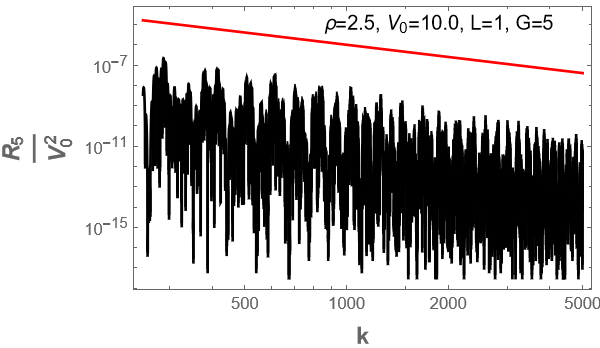} (a) \ \includegraphics[scale=0.56]{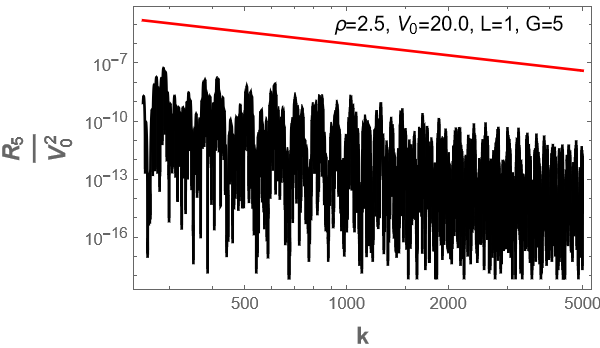} (b)
\caption{\textit{ $\log-\log$ plots showing the scaling behavior of reflection amplitudes $\frac{R_{G}} {V_{0}^{2}}$ for large $k$ for two different values of $V_{0}$. The red curve represent $\frac{1}{k ^{2}}$. It is observed that at large $k$, $R_{G}$ falls off according to $\frac{1}{k ^{2}}$. The potential parameters are shown in the figures.}}
\label{scaling_gsvc}
\end{center}
\end{figure}
In Fig-\ref{rg_gsvc_convergence} we show the behavior of $R_{2}$, $R_{3}$, $R_{4}$ and $R_{5}$ for $\rho=3.5$, $V_{0}=10$ and $L=1$. It is obserbed the $R_{G}$ rapidly converges as $G$ increases. The difference between $R_{4}$ and $R_{5}$ is not visible at nearly all places except at the peaks. This shows the convergence of the product term of Eq. (\ref{vg_zetag}). In Fig-\ref{rg_gsvc_saturation} we show the behavior of $R_{5}$, $R_{10}$ and $R_{15}$ for different $\rho$ values of the potential. The difference between different $R_{G}$ plots (for $G=$5, 10, 15 here) is invisible which again shows the fast convergence of product term of Eq. (\ref{vg_zetag}) with increasing $G$. Similar results are also shown in earlier work on Cantor fractal potential \cite{cantor_f7}. Again, due to the convergence nature of the product term (provided it is evaluated at $V_{G}$) with increasing $G$, it is evident from Eq. (\ref{vg_zetag}) that $R_{G}$ would scale as $\frac{1}{k ^{2}}$ for large $G$ and $k$ values. This is also a proven result for standard Cantor potential \cite{cantor_f7} and we found that this also holds for general SVC potential system. The scaling behavior of $R_{G}$ with $k$ is shown in Fig. \ref{scaling_gsvc} for different potential parameters. 

	\section{Conclusions and Discussions}
	\label{results_discussions}
	Starting from the concept of super periodic potential (SPP), we have demonstrated that a general symmetric  SVC($\rho$)  potential is the special case of SPP with `unit cell' as rectangular barrier. By using the formalism of SPP, we have derived the close form expression of tunneling coefficient $T_{G} (k)$ for an arbitrary stage $G$ of SVC($\rho$)  potential using $q$-Pochhammer symbol. Using the analytical expression $T_{G} (k, \rho)$, we have shown the density plots of $T_{G} (k)$ and 3D plots of reflection coefficient $R_{G} (k)$ for different stages $G$. It is noted that with increasing stage $G$ of SVC($\rho$) potential, the tunneling profile with $k$ shows a saturation with $G$. The reason for this is due to the fact that  consecutively smaller fraction of the remaining previous segments are removed at each stage G. Therefore for higher $G$, only very thin portions are removed from previous segments of stage $G-1$ as compared to the case when $G$ is small. This leads to the saturation of tunneling profile with $k$ for higher $G$.
	
	This system doesn't show the emergence of band like structure which have been noted to occur for periodic rectangular potential with periodicity $N>5$ \cite{griffith, tare}. We have not noted self-similar behaviour of $T_{G} (k)$ with $k$ at present level of our investigation. This can be further direction of our work to find the scaling behavior of $T_{G} (k)$ from this system. There are analytical challenges to this due to analytical series summations of $q$-Pochhammer symbol. It is found that the transmission coefficient from this system show very sharp features of transmission resonances at lower $k$ values which are usually not seen from other hermitian systems. The occurrence of sharp transmission resonances may find applications in the design of transmission filters of very narrow wavelength range.

We have also studied the case of reflection coefficient from this system when the height of the potential changes in a manner that preserve the total area of the potential at each stage $G$. It is shown graphically that  the profile of reflection coefficient rapidly converges with increasing $G$. Further we have shown analytically that for large $k$, the reflection coefficient will vary as $\frac{1}{k^{2}}$. This behavior is also demonstrated graphically.  
\\
\\
{\it \bf{Acknowledgements}}:\\ 
We thank the anonymous referee for the valuable comments which helped us to improve the manuscript. MH acknowledges supports from SPO-ISRO HQ for the encouragement of research activities. One of us (BPM) acknowledges the support from the Research grant under IoE scheme (Number- 6031), UGC-Govt. of India. The present investigation has been carried out under the financial support from BHU RET fellowships to V. N. Singh from Banaras Hindu University (BHU), Varanasi. MU acknowledges the library facility of IIT Delhi.
	
	\newpage
	\begin{center}
		{\large \bf Appendix - A : Derivation of $\zeta_{j}$ }
		\label{Ap_A}
	\end{center}
	\renewcommand{\theequation}{A.\arabic{equation}}
	\setcounter{equation}{0}
	This appendix shows the various steps of calculations to arrive at the general expression of $\zeta_{j}$ used in section \ref{svc_calc} of the paper. If the transfer matrix $M$ 
	\begin{equation}
		M = 
		\begin{pmatrix}
			M_{11} & M_{12} \\
			M_{21} & M_{22}
		\end{pmatrix},
	\end{equation}
	of a `unit cell' potential $V(x)$ defined over a finite interval is known then the transfer matrix of the periodic potential made by periodic repetitions of this `unit cell' is \cite{griffith}, 
	\begin{equation}
		M_{1} = 
		\begin{pmatrix}
			(M_{11})_{1} & (M_{12})_{1} \\
			(M_{21})_{1} & (M_{22})_{1}
		\end{pmatrix}.
	\end{equation}
	Where,
	\begin{equation}
		(M_{11})_{1} =  [M_{11}e^{-iks_{1}} U_{N_{1}-1}(\zeta_{1})-U_{N_{1}-2}(\zeta_{1})]e^{ikN_{1}s_{1}},
		\label{m11_1}
	\end{equation}
	\begin{equation}
		(M_{12})_{1} =  M_{12}U_{N_{1}-1}(\zeta_{1})e^{-ik(N_{1}-1)s_{1}},
		\label{m12_1}
	\end{equation}
	\begin{equation}
		(M_{21})_{1} =  M_{12}^{*}U_{N_{1}-1}(\zeta_{1})e^{ik(N_{1}-1)s_{1}},
		\label{m21_1}
	\end{equation}
	\begin{equation}
		(M_{22})_{1} = [M_{22}e^{iks_{1}}U_{N_{1}-1}(\zeta_{1})-U_{N_{1}-2}(\zeta_{1})]e^{-ikN_{1}s_{1}}.
		\label{m22_1}
	\end{equation}
	In the above, 
	\begin{equation}
		\zeta_{1} = \frac{1}{2}(M_{11}e^{-iks_{1}}+M_{22}e^{iks_{1}}).
		\label{zeta1_expression}
	\end{equation}
	and $U_{N} (y)$ is the Chebyshev polynomial of the second kind. $s_{1}$ is the separation length from the start of one 'unit cell' to the start of another `unit cell'. For Hermitian potentials, the diagonal and off-diagonal elements of the transfer matrix are complex conjugate to each other \cite{griffith}. Thus, $M_{11} = M_{22}^{*}$ and therefore expression for $\zeta_{1}$ can be written as, 
	\begin{equation}
		\zeta_{1}=Re\{M_{22}\}\cos{ks_{1}}-Im\{M_{22}\}\sin{ks_{1}},
		\label{eq5}
	\end{equation}
	where the symbol `Re' and `Im' denotes the real and imaginary parts of a complex number. $\zeta_{1}$ can be further simplified to, 
	\begin{equation}
		\zeta_{1}=\lvert{M_{22}}\rvert \cos({\theta+ks_{1}}).
		\label{zeta1}
	\end{equation}
	Where $\theta$ is the argument of $M_{22}$ i.e., $M_{22}=\lvert{M_{22}}\rvert e^{i\theta}$. For the second order super periodic potential ($j =2$), where the separation between the `unit cell' potential of spatial extent $w_{1}$ is such that $s_{2}$ $\geq$ $w_{1}$ (see Fig-\ref{superperiodicpotential}), the argument of Chebyshev polynomial can be expressed similar to Eq. (\ref{eq5}) as,
	\begin{equation}
		\zeta_{2}=Re\{(M_{22})_{1}\}\cos{ks_{2}}-Im\{(M_{22})_{1}\}\sin{ks_{2}}.
		\label{eq8}
	\end{equation}
	Substituting for $(M_{22})_{1}$ from Eq. (\ref{m22_1}) in the above equation, we arrive at
	\begin{align}
		\begin{split}\label{eq10}
			\zeta_{2} ={}& Re\{M_{22}U_{N_{1}-1}(\zeta_{1})e^{-ik(N_{1}-1)s_{1}}-U_{N_{1}-2}(\zeta_{1})e^{-ikN_{1}s_{1}}\}\cos{ks_{2}}\\ 
			& -Im\{M_{22}U_{N_{1}-1}(\zeta_{1})e^{-ik(N_{1}-1)s_{1}}-U_{N_{1}-2}(\zeta_{1})e^{-ikN_{1}s_{1}}\}\sin{ks_{2}}.
		\end{split}
	\end{align}
	Further, using the expression $M_{22}=\lvert{M_{22}}\rvert e^{i\theta}$ in above, we have
	\begin{align}
		\begin{split} \label{eq11}
			\zeta_{2} ={}& Re\{\lvert{M_{22}}\rvert e^{i\theta}U_{N_{1}-1}(\zeta_{1})e^{-ik(N_{1}-1)s_{1}}-U_{N_{1}-2}(\zeta_{1})e^{-ikN_{1}s_{1}}\}\cos{ks_{2}}\\ 
			& -Im\{\lvert{M_{22}}\rvert e^{i\theta}U_{N_{1}-1}(\zeta_{1})e^{-ik(N_{1}-1)s_{1}}-U_{N_{1}-2}(\zeta_{1})e^{-ikN_{1}s_{1}}\}\sin{ks_{2}} \nonumber 
		\end{split} \\
	\end{align}
	\begin{align}
		\begin{split} \label{eq12}
			= {}& Re\{\lvert{M_{22}}\rvert U_{N_{1}-1}(\zeta_{1})e^{i\{\theta-k(N_{1}-1)s_{1}\}}-U_{N_{1}-2}(\zeta_{1})e^{-ikN_{1}s_{1}}\}\cos{ks_{2}}\\
			& -Im\{\lvert{M_{22}}\rvert U_{N_{1}-1}(\zeta_{1})e^{i\{\theta-k(N_{1}-1)s_{1}\}}-U_{N_{1}-2}(\zeta_{1})e^{-ikN_{1}s_{1}}\}\sin{ks_{2}}.
		\end{split}
	\end{align}
	Using Euler's formula $e^{i\theta}= \cos{\theta}+ i\sin{\theta}$ in the above expression, we have
	\begin{align}
		\begin{split}\label{eq13}
			\zeta_{2}= {}&\big\{\lvert{M_{22}}\rvert U_{N_{1}-1}(\zeta_{1})\cos{\{\theta-k(N_{1}-1)s_{1}\}}-U_{N_{1}-2}(\zeta_{1})\cos{kN_{1}s_{1}}\big\}\cos{ks_{2}}
			\\
			& -\big\{\lvert{M_{22}}\rvert U_{N_{1}-1}(\zeta_{1})\sin{\{\theta-k(N_{1}-1)s_{1}\}}-U_{N_{1}-2}(\zeta_{1})\sin{kN_{1}s_{1}}\big\}\sin{ks_{2}} \nonumber
		\end{split}\\
	\end{align}
	\begin{align}
		\begin{split}
			= {}& \lvert{M_{22}}\rvert U_{N_{1}-1}(\zeta_{1})\cos\{\theta-k(N_{1}-1)s_{1}\}\cos{ks_{2}}-U_{N_{1}-2}(\zeta_{1})\cos(k N_{1}s_{1})\cos({ks_{2}})
			\\
			& -\lvert{M_{22}}\rvert U_{N_{1}-1}(\zeta_{1})\sin{\{\theta-k(N_{1}-1)s_{1}\}}\sin({ks_{2}})+U_{N_{1}-2}(\zeta_{1})\sin({k N_{1}s_{1}})\sin({k s_{2}}) \nonumber
		\end{split}\\
	\end{align}
	\begin{align}
		\begin{split}\label{eq14}
			= {}& \lvert{M_{22}}\rvert U_{N_{1}-1}(\zeta_{1})\Big[\cos{\{\theta-k(N_{1}-1)s_{1}}\}\cos{ks_{2}}-\sin{\{\theta-k(N_{1}-1)s_{1}}\}\sin{ks_{2}}\Big]
			\\
			& -\lvert{M_{22}}\rvert U_{N_{1}-2}(\zeta_{1})\Big[\cos kN_{1}s_{1}\cos ks_{2}+\sin kN_{1}s_{1}\sin ks_{2}\Big].
		\end{split}
	\end{align}
	Using trigonometric identity $\cos{(x\pm y)} = \cos{x}\cos{y} \mp \sin{x}\sin{y}$, and rearranging the above equation, we have the simplified version of $\zeta_{2}$ as
	\begin{equation}
		\zeta_{2}=\lvert{M_{22}}\rvert U_{N_{1}-1}(\zeta_{1})\cos{\big[\theta-k\{(N_{1}-1)s_{1}-s_{2}\}\big]}-U_{N_{1}-2}(\zeta_{1})\cos{\big\{k(N_{1}s_{1}-s_{2})\big\}}.
		\label{z2}
	\end{equation}
	Next, similar to Eq. (\ref{eq8}), $\zeta_{3}$ is expressed through
	\begin{equation}
		\zeta_{3}=Re\{(M_{22})_{2}\}\cos{ks_{3}}-Im\{(M_{22})_{2}\}\sin{ks_{3}},
		\label{eq17}
	\end{equation}
	where $s_{3}$ is the separation between the `unit cell' potential of spatial extent of $w_{2}$ as shown in Fig-\ref{superperiodicpotential}. $(M_{22})_{2}$ is the element of the transfer matrix of super periodic potential of order 2 and is expressed through \cite{griffith, mh_spp}
	\begin{equation}
		(M_{22})_{2}= (M_{22})_{1}e^{-i k (N_{2}-1)s_{2}}U_{N_{2}-1}(\zeta_{2})-U_{N_{2}-2}(\zeta_{2})e^{-i k N_{2}s_{2}}.
		\label{eq18}
	\end{equation}
	Using Eqs. (\ref{m22_1}) and $M_{22}= \vert M_{22} \vert e^{i \theta}$  in above equation, we have
	\begin{align}
		\begin{split} \label{eq19}
			(M_{22})_{2} ={}&M_{22}U_{N_{1}-1}(\zeta_{1})U_{N_{2}-1}(\zeta_{2})e^{-i k[(N_{1}-1)s_{1}+(N_{2}-1)s_{2}]}
			\\
			& -U_{N_{1}-2}(\zeta_{1})U_{N_{2}-1}(\zeta_{2})e^{-i k[N_{1}s_{1}+(N_{2}-1)s_{2}]}
			-U_{N_{2}-2}(\zeta_{2})e^{-i k N_{2}s_{2}}.
		\end{split}
	\end{align}
	Substituting $(M_{22})_{2}$ from Eq. (\ref{eq19}) in Eq. (\ref{eq17}), we have
	\begin{align}
		\begin{split} 
			\zeta_{3}= {}& Re\big\{\lvert{M_{22}}\rvert e^{i\theta}U_{N_{1}-1}(\zeta_{1})U_{N_{2}-1}(\zeta_{2})e^{-ik\{(N_{1}-1)s_{1}+(N_{2}-1)s_{2}\}}
			\\
			& -U_{N_{1}-2}(\zeta_{1})U_{N_{2}-1}(\zeta_{2})e^{-ik\{N_{1}s_{1}+(N_{2}-1)s_{2}\}}-U_{N_{2}-2}(\zeta_{2})e^{-ikN_{2}s_{2}}\big\}\cos{ks_{3}}
			\\
			& -Im\big\{\lvert{M_{22}}\rvert e^{i\theta}U_{N_{1}-1}(\zeta_{1})U_{N_{2}-1}(\zeta_{2})e^{-ik\{(N_{1}-1)s_{1}+(N_{2}-1)s_{2}\}}
			\\
			& -U_{N_{1}-2}(\zeta_{1})U_{N_{2}-1}(\zeta_{2})e^{-ik\{N_{1}s_{1}+(N_{2}-1)s_{2}\}} -U_{N_{2}-2}(\zeta_{2})e^{-ikN_{2}s_{2}}\big\}\sin{ks_{3}} \nonumber
		\end{split}\\
		\begin{split} 
			= {}& Re\big\{\lvert{M_{22}}\rvert U_{N_{1}-1}(\zeta_{1})U_{N_{2}-1}(\zeta_{2})e^{i[\theta-k\{(N_{1}-1)s_{1}+(N_{2}-1)s_{2}\}]}
			\\
			& -U_{N_{1}-2}(\zeta_{1})U_{N_{2}-1}(\zeta_{2})e^{-ik\{N_{1}s_{1}+(N_{2}-1)s_{2}\}}-U_{N_{2}-2}(\zeta_{2})e^{-ikN_{2}s_{2}}\big\}\cos{ks_{3}}
			\\
			& -Im\big\{\lvert{M_{22}}\rvert U_{N_{1}-1}(\zeta_{1})U_{N_{2}-1}(\zeta_{2})e^{i[\theta-k\{(N_{1}-1)s_{1}+(N_{2}-1)s_{2}\}]}
			\\
			& -U_{N_{1}-2}(\zeta_{1})U_{N_{2}-1}(\zeta_{2})e^{-ik\{N_{1}s_{1}+(N_{2}-1)s_{2}\}} -U_{N_{2}-2}(\zeta_{2})e^{-ikN_{2}s_{2}}\big\}\sin{ks_{3}} .
		\end{split}
	\end{align}
	Using Euler's formula in the above expression, we have
	\begin{align}
		\begin{split}\label{eq21}
			\zeta_{3}={}& \lvert{M_{22}}\rvert U_{N_{1}-1}(\zeta_{1})U_{N_{2}-1}(\zeta_{2}) \cos{[\theta-k\{(N_{1}-1)s_{1}+(N_{2}-1)s_{2}\}]}\cos{k s_{3}}
			\\
			&-U_{N_{1}-2}(\zeta_{1})U_{N_{2}-1}(\zeta_{2})\cos{k\{N_{1}s_{1}+(N_{2}-1)s_{2}\}}\cos{k s_{3}}
			\\
			& -U_{N_{2}-2}(\zeta_{2})\cos{(k N_{2}s_{2})}\cos{(k s_{3})}-\lvert{M_{22}}\rvert U_{N_{1}-1}(\zeta_{1})U_{N_{2}-1}(\zeta_{2})
			\\
			& \times \sin{[\theta-k\{(N_{1}-1)s_{1}+(N_{2}-1)s_{2}\}]}\sin{ks_{3}}
			\\
			&+U_{N_{1}-2}(\zeta_{1})U_{N_{2}-1}(\zeta_{2})\sin{k\{N_{1}s_{1}+(N_{2}-1)s_{2}\}}\sin{ks_{3}}
			\\
			& -U_{N_{2}-2}(\zeta_{2})\sin{(kN_{2}s_{2})}\sin{(ks_{3})}.
		\end{split}
	\end{align}
	Rearranging the above equation, we have
	\begin{align}
		\begin{split}\label{eq22}
			\zeta_{3} = {}& \lvert{M_{22}}\rvert U_{N_{1}-1}(\zeta_{1})U_{N_{2}-1}(\zeta_{2})\Big[\cos{[\theta-k\{(N_{1}-1)s_{1}+(N_{2}-1)s_{2}\}]}\cos{k s_{3}}
			\\
			& -\sin{[\theta-k\{(N_{1}-1)s_{1}+(N_{2}-1)s_{2}\}]}\sin{ks_{3}}\Big]
			\\
			& -U_{N_{1}-2}(\zeta_{1})U_{N_{2}-1}(\zeta_{2})\Big[\cos{k\{N_{1}s_{1}+(N_{2}-1)s_{2}\}}\cos{k s_{3}}
			\\
			&-\sin{k\{N_{1}s_{1}+(N_{2}-1)s_{2}\}}\sin{ks_{3}}\Big]
			\\
			&-U_{N_{2}-2}(\zeta_{2})\Big[\cos{(k N_{2}s_{2})}\cos{(k s_{3})}+\sin{(kN_{2}s_{2})}\sin{(ks_{3})}\Big].
		\end{split}
	\end{align}
	Simplification of above equation leads to express $\zeta_{3}$ as
	\begin{align}
		\begin{split}\label{eq23}
			\zeta_{3}= {} &\lvert{M_{22}}\rvert U_{N_{1}-1}(\zeta_{1})U_{N_{2}-1}(\zeta_{2}) \big[\cos{\big\{\theta-k\big((N_{1}-1)s_{1}+(N_{2}-1)s_{2}-s_{3}\big)}\big\}\big]
			\\
			& -U_{N_{1}-2}(\zeta_{1})U_{N_{2}-1}(\zeta_{2})\big[\cos\big\{{k\big(N_{1}s_{1}+(N_{2}-1)s_{2}-s_{3}\big)}\big]
			\\
			& -U_{N_{2}-2}(\zeta_{2})\big[\cos{\{k( N_{2}s_{2}-s_{3}\}}\big].
		\end{split}
	\end{align}
	Next, similar to Eq. (\ref{eq17}), expression of $\zeta_{4}$ for SPP of order-4 is expressed through
	\begin{equation}
		\zeta_{4}=Re\{(M_{22})_{3}\}\cos{ks_{4}}-Im\{(M_{22})_{3}\}\sin{ks_{4}},
		\label{eq25}
	\end{equation}
	where $s_{4}$ is the separation between the `unit cell' potential of spatial extent of $w_{3}$ as shown in Fig-\ref{superperiodicpotential}. $(M_{22})_{3}$ is the element of the transfer matrix of super periodic potential of order-3 and it is expressed through \cite{griffith, mh_spp}
	\begin{equation}
		(M_{22})_{3}= (M_{22})_{2}e^{-i k (N_{3}-1)s_{3}}U_{N_{3}-1}(\zeta_{3})-U_{N_{3}-2}(\zeta_{3})e^{-i k N_{3}s_{3}}.
		\label{eq26}
	\end{equation}
	Substituting $(M_{22})_{2}$ from Eq. (\ref{eq19}) and using $M_{22} = \lvert{M_{22}}\rvert e^{i\theta}$ in above equation, we have
	\begin{align}
		\begin{split}\label{eq27}
			(M_{22})_{3}={}& \lvert{M_{22}}\rvert e^{i\theta}U_{N_{1}-1}(\zeta_{1})U_{N_{2}-1}(\zeta_{2})U_{N_{3}-1}(\zeta_{3})e^{-i k[(N_{1}-1)s_{1}+(N_{2}-1)s_{2}+(N_{3}-1)s_{3}]}
			\\
			& -U_{N_{1}-2}(\zeta_{1})U_{N_{2}-1}(\zeta_{2})U_{N_{3}-1}(\zeta_{3})e^{-i k[N_{1}s_{1}+(N_{2}-1)s_{2}+(N_{3}-1)s_{3}]}
			\\
			&-U_{N_{2}-2}U_{N_{3}-1}(\zeta_{3})e^{-i k\{N_{2}s_{2}+(N_{3}-1)s_{3}\}}-U_{N_{3}-2}(\zeta_{3})e^{-i k N_{3}s_{3}}.
		\end{split}
	\end{align}
	Using above equation in Eq. (\ref{eq25}), $\zeta_{4}$ is expressed through
	\begin{align}
		\begin{split}\label{28}
			\zeta_{4}={}& Re\big\{\lvert{M_{22}}\rvert U_{N_{1}-1}(\zeta_{1})U_{N_{2}-1}(\zeta_{2})U_{N_{3}-1}(\zeta_{3})e^{i\big[\theta- k\{(N_{1}-1)s_{1}+(N_{2}-1)s_{2}+(N_{3}-1)s_{3}\}\big]}
			\\
			&-U_{N_{1}-2}(\zeta_{1})U_{N_{2}-1}(\zeta_{2})U_{N_{3}-1}(\zeta_{3})e^{-i k[N_{1}s_{1}+(N_{2}-1)s_{2}+(N_{3}-1)s_{3}]}
			\\
			&-U_{N_{2}-2}U_{N_{3}-1}(\zeta_{3})e^{-i k\{N_{2}s_{2}+(N_{3}-1)s_{3}\}}-U_{N_{3}-2}(\zeta_{3})e^{-i k N_{3}s_{3}}\big\}\cos{ks_{4}}
			\\
			&-Im\big\{\lvert{M_{22}}\rvert U_{N_{1}-1}(\zeta_{1})U_{N_{2}-1}(\zeta_{2})U_{N_{3}-1}(\zeta_{3})e^{i\big[\theta -k\{(N_{1}-1)s_{1}+(N_{2}-1)s_{2}+(N_{3}-1)s_{3}\}\big]}
			\\
			&-U_{N_{1}-2}(\zeta_{1})U_{N_{2}-1}(\zeta_{2})U_{N_{3}-1}(\zeta_{3})e^{-i k[N_{1}s_{1}+(N_{2}-1)s_{2}+(N_{3}-1)s_{3}]}
			\\
			&-U_{N_{2}-2}U_{N_{3}-1}(\zeta_{3})e^{-i k\{N_{2}s_{2}+(N_{3}-1)s_{3}\}}-U_{N_{3}-2}(\zeta_{3})e^{-i k N_{3}s_{3}}\big\}\sin{ks_{4}}.
		\end{split}
	\end{align}
	Using Euler's formula in the above expression, we have 
	\begin{align}
		\begin{split}\label{eq29}
			\zeta_{4}={}& \lvert{M_{22}}\rvert 
			U_{N_{1}-1}(\zeta_{1})U_{N_{2}-1}(\zeta_{2})U_{N_{3}-1}(\zeta_{3})
			\\
			& \times \cos{\big[\theta- k\{(N_{1}-1)s_{1}+(N_{2}-1)s_{2}+(N_{3}-1)s_{3}\}\big]}\cos{ks_{4}}
			\\
			&-U_{N_{1}-2}(\zeta_{1})U_{N_{2}-1}(\zeta_{2})U_{N_{3}-1}(\zeta_{3})\cos{[ k\{N_{1}s_{1}+(N_{2}-1)s_{2}+(N_{3}-1)s_{3}\}]}\cos{ks_{4}}
			\\
			&-U_{N_{2}-2}U_{N_{3}-1}(\zeta_{3})\cos{[ k\{N_{2}s_{2}+(N_{3}-1)s_{3}\}]}\cos{ks_{4}}
			\\
			&-U_{N_{3}-2}(\zeta_{3})\cos{(k N_{3}s_{3})}\cos{ks_{4}}
			\\
			&-\lvert{M_{22}}\rvert U_{N_{1}-1}(\zeta_{1})U_{N_{2}-1}(\zeta_{2})U_{N_{3}-1}(\zeta_{3})
			\\
			& \times \sin{\big[\theta -k\{(N_{1}-1)s_{1}+(N_{2}-1)s_{2}+(N_{3}-1)s_{3}\}\big]}\sin{ks_{4}}
			\\
			& -U_{N_{1}-2}(\zeta_{1})U_{N_{2}-1}(\zeta_{2})U_{N_{3}-1}(\zeta_{3})\sin{ k\{N_{1}s_{1}+(N_{2}-1)s_{2}+(N_{3}-1)s_{3}\}\big]}\sin{ks_{4}}
			\\
			& -U_{N_{2}-2}U_{N_{3}-1}(\zeta_{3})\sin{ [k\{N_{2}s_{2}+(N_{3}-1)s_{3}\}]}\sin{ks_{4}}
			\\
			&-U_{N_{3}-2}(\zeta_{3})\sin{(k N_{3}s_{3})}\sin{(ks_{4})}.
		\end{split}
	\end{align}
	Rearranging the above equation, we have
	\begin{align}
		\begin{split}\label{eq30}
			\zeta_{4} = {}& \lvert{M_{22}}\rvert 
			U_{N_{1}-1}(\zeta_{1})U_{N_{2}-1}(\zeta_{2})U_{N_{3}-1}(\zeta_{3})
			\\
			&\times \Big[\cos{\big[\theta- k\{(N_{1}-1)s_{1}+(N_{2}-1)s_{2}+(N_{3}-1)s_{3}\}\big]}\cos{ks_{4}}
			\\
			&-\sin{\big[\theta- k\{(N_{1}-1)s_{1}+(N_{2}-1)s_{2}+(N_{3}-1)s_{3}\}\big]}\sin{ks_{4}}\Big]
			\\
			& -U_{N_{1}-2}(\zeta_{1})U_{N_{2}-1}(\zeta_{2})U_{N_{3}-1}(\zeta_{3})
			\\
			&\times \Big[\cos{[ k\{N_{1}s_{1}+(N_{2}-1)s_{2}+(N_{3}-1)s_{3}\}]}\cos{ks_{4}}
			\\
			&+\sin{[ k\{N_{1}s_{1}+(N_{2}-1)s_{2}+(N_{3}-1)s_{3}\}]}\sin{ks_{4}}\Big]
			\\
			&-U_{N_{2}-2}U_{N_{3}-1}(\zeta_{3})\Big[\cos{[ k\{N_{2}s_{2}+(N_{3}-1)s_{3}\}]}\cos{ks_{4}}
			\\
			&+\sin{[ k\{N_{2}s_{2}+(N_{3}-1)s_{3}\}]}\sin{ks_{4}}\Big]
			\\
			&-U_{N_{3}-2}(\zeta_{3})\Big[\cos{(k N_{3}s_{3})}\cos{(ks_{4})}+\sin{(k N_{3}s_{3})}\sin{(ks_{4})}\Big].
		\end{split}
	\end{align}
	Simplification of above equation leads to express $\zeta_{4}$ as
	\begin{align}
		\begin{split}\label{eq31}
			\zeta_{4}={}& \lvert{M_{22}}\rvert 
			U_{N_{1}-1}(\zeta_{1})U_{N_{2}-1}(\zeta_{2})U_{N_{3}-1}(\zeta_{3})
			\\
			&\times \cos{\big[\theta- k\{(N_{1}-1)s_{1}+(N_{2}-1)s_{2}+(N_{3}-1)s_{3}-s_{4}\}\big]}
			\\
			&-U_{N_{1}-2}(\zeta_{1})U_{N_{2}-1}(\zeta_{2})U_{N_{3}-1}(\zeta_{3})
			\\
			&\times \cos{\big[ k\{N_{1}s_{1}+(N_{2}-1)s_{2}+(N_{3}-1)s_{3}-s_{4}\}]}
			\\
			&-U_{N_{2}-2}(\zeta_{2})U_{N_{3}-1}(\zeta_{3})\cos{\big[ k\{N_{2}s_{2}+(N_{3}-1)s_{3}-s_{4}\}]}
			\\
			&-U_{N_{3}-2}(\zeta_{3})\cos{\{k  (N_{3}s_{3}-s_{4})\}}.
		\end{split}
	\end{align}
	Next, similar to Eq. (\ref{eq25}), expression of $\zeta_{5}$ for SPP of order-5 is expressed as, 
	\begin{equation}
		\zeta_{5}=Re\{(M_{22})_{4}\}\cos{ks_{5}}-Im\{(M_{22})_{4}\}\sin{ks_{5}},
		\label{eq32}
	\end{equation}
	where $s_{5}$ is the separation between the starting point of the consecutive `unit cell' potential having spatial extent  $w_{4}$. $(M_{22})_{4}$ is the element of the transfer matrix of super periodic potential of order-4 and is expressed through \cite{griffith, mh_spp}
	\begin{equation}
		(M_{22})_{4}= (M_{22})_{3}e^{-i k (N_{4}-1)s_{4}}U_{N_{4}-1}(\zeta_{4})-U_{N_{4}-2}(\zeta_{4})e^{-i k N_{4}s_{4}}.
		\label{eq33}
	\end{equation}
	Using Eq. (\ref{eq27}) in above equation, $(M_{22})_{4}$ is expressed through
	\begin{align}
		\begin{split}\label{eq34}
			(M_{22})_{4}={}&\lvert{M_{22}}\rvert e^{i\theta}U_{N_{1}-1}(\zeta_{1})U_{N_{2}-1}(\zeta_{2})U_{N_{3}-1}(\zeta_{3})U_{N_{4}-1}(\zeta_{4})
			\\
			&\times e^{-i k[(N_{1}-1)s_{1}+(N_{2}-1)s_{2}+(N_{3}-1)s_{3}+(N_{4}-1)s_{4}]}
			\\
			&-U_{N_{1}-2}(\zeta_{1})U_{N_{2}-1}(\zeta_{2})U_{N_{3}-1}(\zeta_{3})U_{N_{4}-1}(\zeta_{4})
			\\
			&\times e^{-i k[N_{1}s_{1}+(N_{2}-1)s_{2}+(N_{3}-1)s_{3}+(N_{4}-1)s_{4}]}
			\\
			&-U_{N_{2}-2}U_{N_{3}-1}U_{N_{4}-1}(\zeta_{4})e^{-i k\{N_{2}s_{2}+(N_{3}-1)s_{3}+(N_{4}-1)s_{4}\}}
			\\
			&-U_{N_{3}-2}(\zeta_{3})U_{N_{4}-1}(\zeta_{4})e^{-i k\{ N_{3}s_{3}+(N_{4}-1)s_{4}\}}
			\\
			&-U_{N_{4}-2}(\zeta_{4})e^{-i(kN_{4}s_{4})}.
		\end{split}
	\end{align}
	Substituting $(M_{22})_{4}$ from Eq. (\ref{eq34}) in Eq. (\ref{eq32}), we have
	\begin{align}
		\begin{split}\label{eq35}
			\zeta_{5}={}& Re\big\{\lvert{M_{22}}\rvert e^{i\theta}U_{N_{1}-1}(\zeta_{1})U_{N_{2}-1}(\zeta_{2})U_{N_{3}-1}(\zeta_{3})U_{N_{4}-1}(\zeta_{4})
			\\
			&\times e^{i\big[\theta- k\{(N_{1}-1)s_{1}+(N_{2}-1)s_{2}+(N_{3}-1)s_{3}+(N_{4}-1)s_{4}\}\big]}
			\\
			&-U_{N_{1}-2}(\zeta_{1})U_{N_{2}-1}(\zeta_{2})U_{N_{3}-1}(\zeta_{3})U_{N_{4}-1}(\zeta_{4})
			\\
			&\times e^{-i k[N_{1}s_{1}+(N_{2}-1)s_{2}+(N_{3}-1)s_{3}+(N_{4}-1)s_{4}]}
			\\
			&-U_{N_{2}-2}U_{N_{3}-1}(\zeta_{3})U_{N_{4}-1}(\zeta_{4})e^{-i k\{N_{2}s_{2}+(N_{3}-1)s_{3}+(N_{4}-1)s_{4}\}}
			\\
			&-U_{N_{3}-2}(\zeta_{3})U_{N_{4}-1}(\zeta_{4})e^{-i k\{ N_{3}s_{3}+(N_{4}-1)s_{4}\}}
			\\
			&-U_{N_{4}-2}(\zeta_{4})e^{-i(kN_{4}s_{4})}\big\}\cos{ks_{5}}\\
			&-Im\big\{\lvert{M_{22}}\rvert e^{i\theta}U_{N_{1}-1}(\zeta_{1})U_{N_{2}-1}(\zeta_{2})U_{N_{3}-1}(\zeta_{3})U_{N_{4}-1}(\zeta_{4})
			\\
			&\times e^{i\big[\theta- k\{(N_{1}-1)s_{1}+(N_{2}-1)s_{2}+(N_{3}-1)s_{3}+(N_{4}-1)s_{4}\}\big]}
			\\
			&-U_{N_{1}-2}(\zeta_{1})U_{N_{2}-1}(\zeta_{2})U_{N_{3}-1}(\zeta_{3})U_{N_{4}-1}(\zeta_{4})
			\\
			&\times e^{-i k[N_{1}s_{1}+(N_{2}-1)s_{2}+(N_{3}-1)s_{3}+(N_{4}-1)s_{4}]}
			\\
			&-U_{N_{2}-2}U_{N_{3}-1}(\zeta_{3})U_{N_{4}-1}(\zeta_{4})e^{-i k\{N_{2}s_{2}+(N_{3}-1)s_{3}+(N_{4}-1)s_{4}\}}
			\\
			&-U_{N_{3}-2}(\zeta_{3})U_{N_{4}-1}(\zeta_{4})e^{-i k\{ N_{3}s_{3}+(N_{4}-1)s_{4}\}}
			\\
			&-U_{N_{4}-2}(\zeta_{4})e^{-i(kN_{4}s_{4})}\big\}\sin{ks_{5}}.
		\end{split}
	\end{align}
	Using Euler's formula in above expression, we have
	\begin{align}
		\begin{split}\label{36}
			\zeta_{5}={}&\lvert{M_{22}}\rvert 
			U_{N_{1}-1}(\zeta_{1})U_{N_{2}-1}(\zeta_{2})U_{N_{3}-1}(\zeta_{3})U_{N_{4}-1}(\zeta_{4})
			\\
			&\times \cos{\big[\theta- k\{(N_{1}-1)s_{1}+(N_{2}-1)s_{2}+(N_{3}-1)s_{3}+(N_{4}-1)s_{4}\}\big]}\cos{ks_{5}}
			\\
			&-U_{N_{1}-2}(\zeta_{1})U_{N_{2}-1}(\zeta_{2})U_{N_{3}-1}(\zeta_{3})U_{N_{4}-1}(\zeta_{4})
			\\
			&\times \cos{[ k\{N_{1}s_{1}+(N_{2}-1)s_{2}+(N_{3}-1)s_{3}+(N_{4}-1)s_{4}\}]}\cos{ks_{5}}
			\\
			&-U_{N_{2}-2}U_{N_{3}-1}(\zeta_{3})U_{N_{4}-1}(\zeta_{4})\cos{[ k\{N_{2}s_{2}+(N_{3}-1)s_{3}+(N_{4}-1)s_{4}\}]}\cos{ks_{5}}
			\\
			&-U_{N_{3}-2}(\zeta_{3})U_{N_{4}-1}(\zeta_{4})\cos{\big[k (N_{3}s_{3}+(N_{4}-1)s_{4})\big]}\cos{ks_{5}}
			\\
			&-U_{N_{4}-2}(\zeta_{4})\cos{(kN_{4}s_{4})}\cos{(ks_{5})}
			\\
			&-\lvert{M_{22}}\rvert U_{N_{1}-1}(\zeta_{1})U_{N_{2}-1}(\zeta_{2})U_{N_{3}-1}(\zeta_{3})U_{N_{4}-1}(\zeta_{4})
			\\
			&\times \sin{\big[\theta -k\{(N_{1}-1)s_{1}+(N_{2}-1)s_{2}+(N_{3}-1)s_{3}+(N_{4}-1)s_{4}\}\big]}\sin{ks_{5}}
			\\
			&-U_{N_{1}-2}(\zeta_{1})U_{N_{2}-1}(\zeta_{2})U_{N_{3}-1}(\zeta_{3})U_{N_{4}-1}(\zeta_{4})
			\\
			&\times \sin{\big[ k\{N_{1}s_{1}+(N_{2}-1)s_{2}+(N_{3}-1)s_{3}+(N_{4}-1)s_{4}\}\big]}\sin{ks_{5}}
			\\
			&
			-U_{N_{2}-2}U_{N_{3}-1}(\zeta_{3})U_{N_{4}-1}(\zeta_{4})\sin{ [k\{N_{2}s_{2}+(N_{3}-1)s_{3}+(N_{4}-1)s_{4}\}]}\sin{ks_{5}}
			\\
			&-U_{N_{3}-2}(\zeta_{3})U_{N_{4}-1}(\zeta_{4})\sin{\big[k (N_{3}s_{3}+(N_{4}-1)s_{4})\big]}\sin{(ks_{5})}
			\\
			&-U_{N_{4}-2}(\zeta_{4})\sin{(kN_{4}s_{4})}\sin{(ks_{5})}.
		\end{split}
	\end{align}
	Rearranging the above equation, we have
	\begin{align}
		\begin{split}\label{eq37}
			\zeta_{5}={}&\lvert{M_{22}}\rvert 
			U_{N_{1}-1}(\zeta_{1})U_{N_{2}-1}(\zeta_{2})U_{N_{3}-1}(\zeta_{3})U_{N_{4}-1}(\zeta_{4})
			\\
			&\times \Big[\cos{\big[\theta- k\{(N_{1}-1)s_{1}+(N_{2}-1)s_{2}+(N_{3}-1)s_{3}+(N_{4}-1)s_{4}\}\big]}\cos{ks_{5}}
			\\
			&-\sin{\big[\theta- k\{(N_{1}-1)s_{1}+(N_{2}-1)s_{2}+(N_{3}-1)s_{3}+(N_{4}-1)s_{4}\}\big]}\sin{ks_{5}}\Big]
			\\
			& -U_{N_{1}-2}(\zeta_{1})U_{N_{2}-1}(\zeta_{2})U_{N_{3}-1}(\zeta_{3})U_{N_{4}-1}(\zeta_{4})
			\\
			&\times \Big[\cos{[ k\{N_{1}s_{1}+(N_{2}-1)s_{2}+(N_{3}-1)s_{3}+(N_{4}-1)s_{4}\}]}\cos{ks_{5}}
			\\
			&+\sin{[ k\{N_{1}s_{1}+(N_{2}-1)s_{2}+(N_{3}-1)s_{3}+(N_{4}-1)s_{4}\}]}\sin{ks_{5}}\Big]
			\\
			&-U_{N_{2}-2}U_{N_{3}-1}(\zeta_{3})U_{N_{4}-1}(\zeta_{4})
			\\
			&\times \Big[\cos{[ k\{N_{2}s_{2}+(N_{3}-1)s_{3}+(N_{4}-1)s_{4}\}]}\cos{ks_{5}}
			\\
			&+\sin{[ k\{N_{2}s_{2}+(N_{3}-1)s_{3}+(N_{4}-1)s_{4}\}]}\sin{ks_{5}}\Big]
			\\
			&-U_{N_{3}-2}(\zeta_{3})U_{N_{4}-1}(\zeta_{4})
			\\
			&\times \Big[\cos{\big[k (N_{3}s_{3}+(N_{4}-1)s_{4})\big]}\cos{ks_{5}}+\sin{\big[k (N_{3}s_{3}+(N_{4}-1)s_{4})\big]}\sin{ks_{5}}\Big]
			\\
			&-U_{N_{4}-2}(\zeta_{4})\Big[\cos{(kN_{4}s_{4})}\cos{(ks_{5})}+\sin{(kN_{4}s_{4})}\sin{(ks_{5})}\Big].
		\end{split}
	\end{align}
	Simplification of above equation leads to express $\zeta_{5}$ as
	\begin{equation}
		\begin{split}\label{eq38}
			\zeta_{5}={}&\lvert{M_{22}}\rvert 
			U_{N_{1}-1}(\zeta_{1})U_{N_{2}-1}(\zeta_{2})U_{N_{3}-1}(\zeta_{3})U_{N_{4}-1}(\zeta_{4})
			\\
			&\times \cos{\big[\theta- k\{(N_{1}-1)s_{1}+(N_{2}-1)s_{2}+(N_{3}-1)s_{3}+(N_{4}-1)s_{4}-s_{5}\}\big]}
			\\
			&-U_{N_{1}-2}(\zeta_{1})U_{N_{2}-1}(\zeta_{2})U_{N_{3}-1}(\zeta_{3})U_{N_{4}-1}(\zeta_{4})
			\\
			& \times \cos{\big[ k\{N_{1}s_{1}+(N_{2}-1)s_{2}+(N_{3}-1)s_{3}+(N_{4}-1)s_{4}-s_{5}\}]}
			\\
			&-U_{N_{2}-2}(\zeta_{2})U_{N_{3}-1}(\zeta_{3})U_{N_{4}-1}(\zeta_{4})
			\\
			& \times \cos{\big[ k\{N_{2}s_{2}+(N_{3}-1)s_{3}+(N_{4}-1)s_{4}-s_{5}\}]}
			\\
			&-U_{N_{3}-2}(\zeta_{3})U_{N_{4}-1}(\zeta_{4})\cos{\{k  (N_{3}s_{3}+(N_{4}-1)s_{4}-s_{5})\}}
			\\
			&-U_{N_{4}-2}(\zeta_{4})\cos{\{k(N_{4}s_{4}-s_{5}) \}}.
		\end{split}
	\end{equation}
	In the similar fashion, for  SPP of order-$j$ we have the expression of $\zeta_{j}$ as,
	\begin{equation}
		\zeta_{j}=  Re\{(M_{22})_{j-1}\}\cos{ks_{j}}-Im\{(M_{22})_{j-1}\}\sin{ks_{j}}.
		\label{eq39}
	\end{equation}
	Following the similar procedures as outlined previously, one can calculate the analytical expressions for $\zeta_{6}, \zeta_{7}, \zeta_{8},...,\zeta_{j}$. It is easy to see that the above expressions derived for various $\zeta_{j}$ can be expressed through the following series expression,
	\begin{multline}
		\zeta_{j}= \lvert{M_{22}}\rvert \cos{\Bigg[ \theta-k\Bigg\{\sum_{p=1}^{j-1}(N_{p}-1)s_{p}-s_{j}\Bigg\}}\Bigg]\prod_{p=1}^{j-1}U_{N_{p}-1}(\zeta_{p})\\-\sum_{r=1}^{j-2} \cos{\Bigg[k\Bigg\{\sum_{p=r}^{j-1}N_{p}s_{p}-\sum_{p=r+1}^{j}s_{p})\Bigg\}\Bigg]}U_{N_{r}-2}(\zeta_{r})\prod_{p=r+1}^{j-1}U_{N_{p}-1}(\zeta_{p})\\
		-U_{N_{j-1}-2}(\zeta_{j-1})\cos{k(N_{j-1}s_{j-1}-s_{j})},
		\label{40}
	\end{multline}
	Above series representation of $\zeta_{j} \ \forall j \in \{3,4,..,G \} $ given by Eq. (\ref{40}) and expression for $\zeta_{1}$ and $\zeta_{2}$ are expressed by using Eq. $(\ref{zeta1})$ and $(\ref{z2})$.

\end{document}